\documentclass[traditabstract]{aa} % for a referee version
\usepackage{graphicx}
\usepackage{txfonts}
\usepackage{rotating}
\usepackage{multirow} 

\usepackage{appendix}
\usepackage{latexsym}
\usepackage{lscape}
\usepackage{afterpage}

\begin{document}
%
%   \title{VLT VIMOS and SINFONI integral field spectroscopy of the multiphase ISM and the stars in the Luminous InfraRed Galaxy F11506-3851 (ESO 320-G030)}
%\subtitle{Spatially resolved kinematics and galactic wind}

  \title{Spatially resolved kinematics, galactic wind, and quenching of star formation in the luminous infrared galaxy IRAS F11506-3851}

     \author{S.~Cazzoli\inst{\ref{inst1}}\and S.~Arribas \inst{\ref{inst1}}\and L.~Colina\inst{\ref{inst1}}\and J.~Piqueras-L\'opez\inst{\ref{inst1}}\and E.~Bellocchi\inst{\ref{inst1}}\and B.~Emonts\inst{\ref{inst1}}\and R.~Maiolino\inst{\ref{inst2},\ref{inst3}}}  %\and
          %\fnmsep\thanks{Just to show the usage
          %of the elements in the author field}
   \institute{CSIC - Departamento de Astrofisica-Centro de Astrobiologia (CSIC-INTA),Torrejon de Ardoz, Madrid, Spain \\ \email{scazzoli@cab.inta-csic.es}\label{inst1} \and Cavendish Laboratory, University of Cambridge 19 J. J. Thomson Avenue, Cambridge CB3 0HE, UK\label{inst2} \and Kavli Institute for Cosmology, University of Cambridge, Madingley Road, Cambridge CB3 0HA, UK\label{inst3}
}

   \date{Received  20 December 2013 / Accepted 11 May 2014 }

% \abstract{}{}{}{}{} 
% 5 {} token are mandatory
 
  \abstract
{We present a multi-wavelength integral field spectroscopic (IFS) study of the low-z luminous infrared galaxy IRAS F11506-3851  (ESO 320-G030) on the basis of the moderate spectral resolution observations (R\,$\sim$\,3400\,-\,4000) taken with the VIMOS and SINFONI instruments at the ESO VLT. The morphology and the 2D kinematics of the gaseous (neutral and ionized) and stellar components have been mapped in the central regions ($<\,$\,3 kpc) using the NaD$\lambda$$\lambda$5890,5896 $\AA$ absorption doublet, the H$\alpha$$\lambda$6563  $\AA$ line, and the near-IR CO(2-0)$\lambda$2.293\,$\mu$m and CO(3-1)$\lambda$2.322\,$\mu$m bands.  \\
The kinematics of the ionized gas and the stars are dominated by rotation, with large observed velocity amplitudes ($\Delta$V(H$\alpha$)\,=\,203\,$\pm$\,4 km\,s$^{-1}$; $\Delta$V(CO)\,=\,188 $\pm$ 11 km\,s$^{-1}$, respectively) and centrally peaked velocity dispersion maps ($\sigma_{\rm c}$(H$\alpha$)\,=\,95\,$\pm$\,4 km\,s$^{-1}$ and $\sigma_{\rm c}$(CO)\,=\,136 $\pm$ 20 km\,s$^{-1}$). The stars lag behind the warm gas and represent a dynamically hotter system, as indicated by the observed V/$\sigma$  ratios (4.5 and 2.4 for the gas and the stars, respectively). Thanks to these IFS data we have disentangled the contribution of the stars and the interstellar medium to the NaD feature, finding that it is dominated by absorption of neutral gas clouds in the ISM ($\sim$\,2/3 of total EW).  The 2D kinematics of the neutral gas shows a complex structure dominated by two main components. On the one hand, the thick slowly rotating disk ($\Delta$V(NaD)\,=\,81\,$\pm$\,12 km\,s$^{-1}$) lags significantly compared to the ionized gas and the stars, and it has an irregular and off-center velocity dispersion map (with values of up to $\sim$\,150 km\,s$^{-1}$ at $\sim$\,1 kpc from the nucleus). On the other hand, a kpc-scale neutral gas outflow perpendicular to the disk, as  is revealed by the large blueshifted velocities (in the range 30\,-\,154 km\,s$^{-1}$) observed along the galaxy's semi-minor axis (within the inner 1.4 kpc).\\  % covering a triangular region of $\sim$ 0.9 $kpc^{2}$.
On the basis of a simple free wind scenario, we derive an outflowing mass rate ($\dot{M_{\rm w}}$) in neutral gas of about 48 M$_{\rm \odot}$ yr$^{-1}$.  Although this implies a global mass loading factor (i.e.,  $\eta$\,=\,$\dot{M_{\rm w}}$/SFR) of  $\sim$\,1.4, the 2D distribution of the ongoing SF as traced by the H$\alpha$ emission map suggests a much larger value of $\eta$ associated with the inner regions (R\,$<$\,200 pc), where the current observed star formation (SF) represents only $\sim$\,3 percent of the total. However, the relatively strong emission by SNe in the central regions, as traced by the [FeII] emission, indicates recent strong episodes of SF. Therefore, our data show clear evidence of the presence of a strong outflow with large loading factors associated with the nuclear regions, where recent starburst activity took place about 7 Myr ago, although it currently shows relatively modest SF levels. All together these results strongly suggest that we are witnessing (nuclear) quenching due to SF feedback in IRAS F11506-3851. However, the relatively large mass of molecular gas detected in the nuclear region via the H$_2$ 1$-0$S(1) line suggests that further episodes of SF may take place again.} 

\keywords{galaxies: starburst ---   ISM: jets and  outflows --- ISM: kinematics and dynamics}
\titlerunning{Spatially resolved kinematics and galactic wind in IRAS F11506-3851}
\authorrunning {Cazzoli et al.}

\maketitle
%
%________________________________________________________________

%The key point of studying feedbacks is to estimate the contribution of mass, energy and metals from  GWs  back to their environments over a wide scale (e.g., $\sim$ 1-10 kpc, \cite{Harrison}).\\
%By investigating the neutral gas kinematics and properties we  constrain the outflow parameters and thus deepen the mechanism driving gas outside  galaxies and on  the relationship between properties of GWs and their hosts.\\

\newcommand{\Ha}{H$\alpha$}
\newcommand{\kms}{$\mbox{kms}^{-1}$}
\newcommand{\mg}{H$_{\rm 2}$1-0S} %mg= molecular gas
\newcommand{\co}{CO(2-0)}
\newcommand{\obj}{IRAS F11506-3851}
\newcommand{\Piquerasprep}{Piqueras L\'opez et al. in prep}
%?\newcommand{\Arribasprep}{Arribas et al. in prep}

%On the basis of a simple free wind scenario, we derive an outflowing mass rate ($\dot{M_{\rm w}}$) in neutral gas of about 48 M$_{\rm \odot}$ yr$^{-1}$.  Although this implies a global mass loading factor (i.e.,  $\eta$ = $\dot{M_{\rm w}}$/SFR) of  $\sim$ 1.4, the 2D distribution of the ongoing SF as traced by the H$\alpha$ emission map suggests a much larger value of $\eta$ associated to the inner regions (r$<$ 200 pc), where the current observed SF represents only $\sim$ 3 percent of the total. However, the relative strong emission by SNe in the central regions, as traced by the [FeII] emission, indicates recent strong episodes of SF. Therefore, our data show clear evidence for the presence of a strong outflow with large loading factors associated to the nuclear regions, where recent starburst activity has taken place about 7 Myr ago, though it currently shows relatively modest SF levels. Taking all together, these results strongly suggest that we are witnessing quenching due to SF feedback in IRAS F11506-3851. However, the relatively large mass of molecular gas detected in the nuclear region via the H$_2$ 1$-0$S(1) line suggests that future (recurrent ?) episodes of SF may take place again. 

\section{Introduction}

Galactic winds (GWs) are believed to play an important role in the evolution of galaxies. They could regulate and quench both the star formation and the black hole activity, being  also the primary mechanism by which dust and metals are redistributed over large scales in the interstellar medium (ISM), or even expelled outside the galaxy into the intergalactic medium  (IGM; e.g., \cite{Rwind}).  Cosmological models of galaxy evolution require energetic outflows to reproduce the observed properties of galaxies -- i.e.,  without a strong (stellar or AGN) feedback, they lead  galaxies to have much higher star forming rates (SFRs) and larger stellar masses than observed (\cite{Hopkins2012}).\\
From an observational point of view,  GWs are found at all cosmic epochs. During the peak of star formation (z$\sim$2-3)  outflows are very prominent and ubiquitous (\cite{Erb2012}; \cite{Maiolino2012}; \cite{Martin2012}; \cite{CanoDiaz2012}).  In the local universe, all the star-forming galaxies in which the star formation rate per unit area exceeds $\Sigma_{\rm SFR}$ $\geq$ $10^{-1}$ $M_{\rm \sun}$ $yr^{-1}$ $kpc^{-2}$ seem to develop GWs (\cite{H02}; \cite{Lehnert96}).  However, despite the theoretical need for outflows and the strong evidence of their existence, it is still unclear if their observed properties (e.g., velocities, mass loading rates) are consistent with those required by current models to quench the star formation   and/or  to expel significant amounts of baryons from the host galaxy  (e.g.,   \cite{Hopkins2012}; \cite{Arribas2014}).\\
Local luminous and ultra-luminous infrared galaxies [U]LIRGs (LIRGs, $L_{\rm IR}$\,=\,$L_{\rm (8-1000 \mu m)}$\,=\,$10^{11}$-$10^{12}$\,$L_{\rm \sun}$, ULIRGs, $L_{\rm IR}$\,$\geq$ $10^{12}$, respectively) are interesting  populations for the study of outflows. On the one hand these objects show the most conspicuous cases for outflows in the local universe (\cite{H00}; \cite{R02}; \cite{M05}; \cite{R05b}, c).  On the other hand,  local [U]LIRGs have a starburst activity similar to that found for normal (i.e., main-sequence; Elbaz et al. 2012)  high-z star forming galaxies (SFGs), and share with distant galaxies some other basic structural (\cite{Arribas2012}) and kinematical (\cite{Bellocchi2013}) properties. Therefore, although the actual analogy between local and high-z [U]LIRGs is under discussion, local [U]LIRGs offer the opportunity to study the outflow phenomenon at environments similar to that observed at high-z, but with a much higher S/N and spatial resolution. More specifically,  local LIRGs may be particularly relevant in the comparison of local and distant outflows, as some authors have suggested that high-z [U]LIRGs are scaled-up versions of lower luminosity local [U]LIRGs (e.g.,  \cite{Takagi2010};  \cite{Muzzin2010}).\\
Most of the GWs studies in local [U]LIRGs have been based on optical spectral features. Emission lines (mainly H$\alpha$) allowed the characterization of  the warm ionized gas outflows (e.g., \cite{H90}; \cite{Soto2012} and references there in), while the  cold neutral gas entraining in the winds is detected through absorption features like the sodium doublet, NaD $\lambda$$\lambda$5890,5896 (e.g., \cite{H00}; \cite{SM04};  \cite{M05}; \cite{R05a}, b, c; \cite{Chen10}).  The large majority of these studies are based on long-slit spectroscopic observations, and, despite its obvious advantages when studying an anisotropic phenomenon (e.g., outflows) in structurally complex objects (e.g., [U]LIRGs), the use of integral field spectroscopic (IFS) techniques for these studies has been rather limited. Initial works were focused on ionized outflows in extreme sources, likely not very representative of the [U]LIRGs population at large (Mrk273:  \cite{Colina1999}, Arp220:  \cite{Arribas2001}).  Ionized outflow IFS-studies for larger and more representative samples of [U]LIRGs have only recently been carried out (e.g.,  \cite{Westmoquette2012}; \cite{Bellocchi2013}; Arribas et al. 2014, and references there in).  
Works using IFS techinique has been recently expanded to study the multi-phase (including neutral and molecular gas) structure of outflows, mainly focused in the most luminous merger systems (ULIRGs;  \cite{R11}, \cite{RV13}).  These studies have shown that in these objects outflows are massive enough to provide negative feedback to star formation. The neutral outflows in the potentially interesting LIRG range have been explored less, although at even lower luminosities it is worth mentioning the works by  \cite{Vicente07}, \cite{Westmoquette2012},  and \cite{Davis12}.  \\       
In this paper we discuss the multi-phase/component structure and wind properties of the nearby LIRG   \obj \, (ESO 320-G030). It is based on  IFS observation with  VIMOS and SINFONI at VLT, thanks to which we have obtained the 2D distribution of the of the NaD absorption doublet, the \Ha \, emission line, and the  \co \,$\lambda$2.322$\mu$m\, absorption band profiles over its inner region (within R $\leq$ 3 kpc). This allowed us to trace the structural and kinematical properties of different galaxy components (gas and stars) and gaseous phases (ionized and neutral).  The source \obj \  is a regular isolated disk (Fig.\,\ref{immagine}) whose main basic properties are summarized in Table\,\ref{source}. It hosts a weak AGN that has a minor contribution to the total infrared luminosity ($<$\,4$\%$ \cite{PereiraSantaella2010}). The mean velocity dispersion in the disk of \obj\  ($\sigma$(H$\alpha$) $\sim$ 40 km\,s$^{-1}$; \cite{Bellocchi2013}) suggests a dynamical status of the ISM intermediate between that found for high-z populations (e.g.,  $\sigma$ $\sim$ 60-90  km\,s$^{-1}$;  \cite{ForsterS2011}) and that of local spirals ($\sigma$  $\sim$ 15-30  km\,s$^{-1}$,  \cite{Epinat2010}). \\
The structure of this paper is as follows. In Sect.\, 2, we briefly describe the observations and data reduction.  In Sect.\,3 we present the data analysis, including details about the spectra fitting for the different features used. Section\,4 is dedicated to describing and discussing the results. The 2D properties of the different galaxy components are commented and discussed, with special focus on the characteristics of the galactic wind and its relation with other galaxy properties. Section 5 summarizes the main results.  We assume $H_{\rm 0}$\,=\,70 km $s^{-1}$Mpc$^{-1}$ and the standard $\Omega_{\rm m}$\,=\,0.3,  $\Omega_{\rm \Lambda}$\,=\,0.7 cosmology.

%It has no traces of hosting a powerful AGN (\cite{PereiraSantaella2010}) -- \textbf{MVM: It host an AGN, but it is not the dominant contribution to the IR-Luminosity}. 

\label{intro}

\section{Observations and data reduction}

\label{Observations}

	\subsection{VLT-VIMOS and SINFONI observations}

The optical data analyzed in this paper were obtained using the integral field unit (IFU) of VIMOS (\cite{LeFevre03}) at the Very Large Telescope (VLT). Details of these VIMOS observations are given in Rodr\'iguez-Zaur\'in et al. (2010 and references therein), with only basic information provided here. The VIMOS-IFU has a 40\,$\times$\,40 fiber array and, therefore, it allows us to obtain a total of 1600 simultaneous spectra per pointing. We used the HR Orange grism) covering the 5250-7400 $\AA$ spectral range with a spectral  resolution of R $\sim$\,3400 (dispersion of 0.6 $\AA$$pix^{-1}$) over a field of view (FoV) of  27$\arcmin$$\arcmin$\,$\times$\,27$\arcmin$$\arcmin$. During the observations of  \obj \,  the average seeing was $\sim$ 0.9\rq\rq.\\
We also made use of near-IR SINFONI  observations of the \co \, absorption bands, which were carried out in the  K band (1.95-2.45 $\mu$m) with the 250 mas scale configuration. The spectral resolution and FoV for this configuration are R $\sim$ 4000 (dispersion of 2.45 $\AA$$pix^{-1}$) and $\sim$ 8$\arcmin$$\arcmin$ $\times$ 8$\arcmin$$\arcmin$ (equivalent to 68 $\times$ 72  spaxels in the  data cube), respectively. Further details about the near-IR observations of this object are given in \cite{Piqueras12}.

\label{obs}
	
				 	\begin{figure}%[b]
		   \centering
	 	       \includegraphics[scale=0.48]{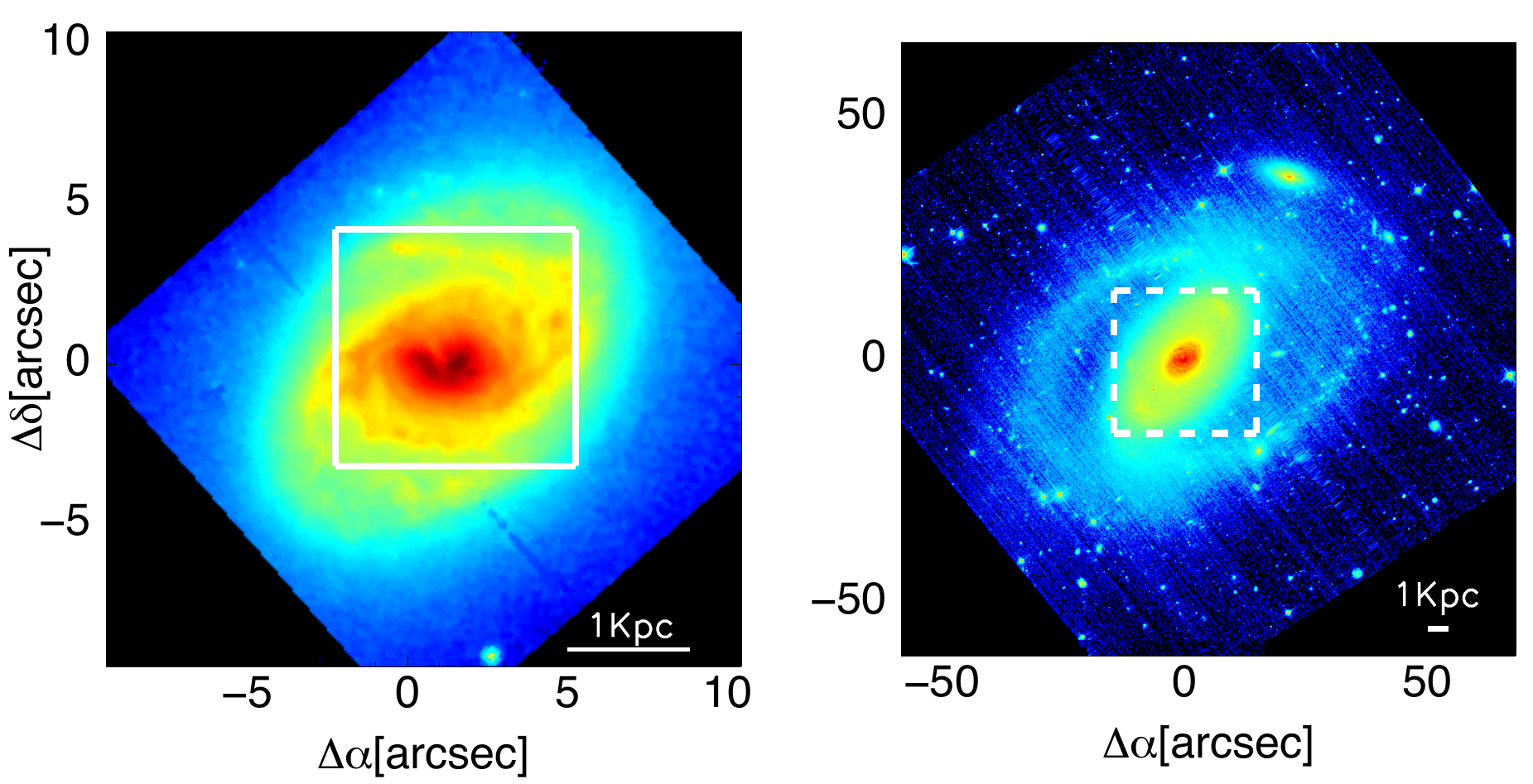}	      %				 
				  \caption{\textit{Hubble Space Telescope (HST)} images of  \obj. The left panel shows the \textit{HST}-NICMOS (Near-Infrared Camera and Multi-Object Spectrometer) F160W image (\cite{AH06}), whereas the right panel displays the  \textit{HST}-ACS (Advanced Camera for Surveys) F814W image  (\cite{Robishaw}). Overlaid are the  SINFONI  (9$\arcmin$$\arcmin$ $\times$ 9$\arcmin$$\arcmin$, left panel) and VIMOS (27$\arcmin$$\arcmin$ $\times$ 27$\arcmin$$\arcmin$, right panel) FoVs (the latter shown with  dashed lines). We note that the FoV covered by the NaD absorption (20$\arcmin$$\arcmin$ $\times$  17$\arcmin$$\arcmin$) cover only the central region of the VIMOS FoV. See text for details. The white bar shows a linear scale of 1 kpc (corresponding to 4$\farcs$53 at the adopted distance, Table \ref{source}). }	
 \label{immagine}    %Spiral arms and several knots of star formation are clearly visible.
		\end{figure}

		\begin{table}
			\caption[Subsample]{General properties of  \obj}
				\begin{center}
					\tiny{\begin{tabular}{l  c  c}
					\hline
					\hline
						Properties & value &  References\\
				\hline
					z       &  0.01078 & \cite{DeVacouler91}\\	
					$D_{\rm L}$ [Mpc]   & 46.6$^a$  & \multirow{2}{*}{}\\
					scale  [pc/arcsec]   &  221$^a$   & \\
					log($L_{\rm IR}$/$L_{\rm \sun}$) & 11.30$^b$ &\cite{Moshir} \\	
						SFR  [$M_{\rm \sun}$/$yr^{-1}$]& 34$^c$ & \cite{RZ10} \\
					$M_{\rm dyn}$ [$M_{\rm \sun}$]  &  (5  $\pm$ 1)  $\times$  $10^{10}$ & \cite{Bellocchi2013}\\
					 Inclination  angle & (37 $\pm$ 3)$\degr$  & \cite{Bellocchi2013}\\
					Nuclear Spectral Class. & HII Galaxy& \cite{Broek91} \\				
					% Morphology Class  & Isolated disk &  \\   
					\hline
					\hline
					\end{tabular}
					\label{source}}
					\end{center} 
	\tiny{Notes. --- $^a$ Luminosity distance and scale using Edward L. Wright Cosmology calculator (\cite{Wright2006}). $^b$ Infrared luminosity (in solar bolometric luminosity and logarithmic units) defined as $L_{\rm IR}$\,=\,$L_{\rm (8-1000 \mu m)}$ and calculated using the fluxes in the four IRAS bands as given in \cite{Sanders2003}.  $^c$ Star formation rate computed  from the IR luminosity (integrated over the range 8-1000 $\mu$m) following \cite{Kennicutt1998}.} %\textbf{$^d$ Inclination  angle defined as the ratio between the major and minor axes of the object (\cite{Bellocchi2013}).} 		
		 	\end{table}
%
					
%associated to each pointing

	\subsection{Data reduction}  

Raw VIMOS data were reduced with a combination of the pipeline provided by ESO via ESOREX  (version 3.6.1 and 3.6.5) including sky and bias subtraction, flat field correction, spectra tracing and extraction, correction of fiber and pixel transmission,  the flux and wavelength calibration flux, and  a set of customized IDL and IRAF scripts. Once the  four  quadrants were  individually reduced (for the four dithered positions), they were combined into a single data-cube  consisting  of 44\,$\times$\,44 spaxels (i.e., 1936 spectra). The instrumental profile correction and the accuracy of wavelength calibration were measured using the [O I] 6300.3 $\AA$ and  the 5577 $\AA$  night-sky lines. The effect of instrumental dispersion, i.e., $\sigma_{\rm instr}$, (0.75\,$\pm$\,0.17) $\AA$ was corrected for by subtracting it in quadrature from the observed line dispersion. \\
The basic reduction of the SINFONI observations was also accomplished via ESOREX  (version 2.0.5).   Details of this process, including the flux calibration method are presented in \cite{Piqueras12}. In brief, the observed object and sky frames were corrected from dark subtraction, flat fielding, detector linearity, geometrical distortion, and  wavelength calibration.  The background sky emission was removed using the method outlined in \cite{Davies2007}. A set of IDL routines were used to perform the flux calibration on the cube while taking the relative shifts in the jittering pattern into account, before being merged in the final datacube.

\label{DR}
\section{Data analysis}
\label{DA}

 \subsection{Line fitting}
 
 \subsubsection{NaD absorption line profiles}

Most of the NaD-absorption features in \obj \, show smooth and unblended line profiles and therefore they were modeled with a single kinematic component, i.e., a Gaussian pair (Fig.\,\ref{spaxels}, upper-left panel).  However some spectra, mainly located in the central region, have asymmetrical  and complex profiles that suggest the presence of an additional secondary component, and therefore they require two Guassian pairs for the fit (Fig.\,\ref{spaxels}, bottom-left panel).\\ 
The existence of two global components is expected as the NaD absorption feature may have originated in the neutral gas clouds of the ISM as well as in the stars.  After experimenting  with these two components it was clear that the main component is due to absorption by neutral gas, while the secondary component has a stellar origin (see Sect.\,\ref{stellarNaD} for a detailed discussion of their properties and respective roles). Fitting the neutral gas absorption with a Gaussian pair represents a simplification of the real case as many individual clouds (i.e., subcomponents) are likely to be present along the line of sight (\cite{R05a}). However, this approach allows us to characterize the global neutral gas  kinematics without introducing other model dependent variables. An alternative method that allows more complex intensity profiles is described by \cite{R05a}.\\ 
The observed absorption profiles of the individual spectra are fitted with a Levenberg-Marquardt least-squares fitting routine (MPFITEXPR, implemented by \cite{Markwardt}) in the IDL environment. For the main (neutral gas) component the central wavelength is a free parameter, while the widths are constrained to be equal for the two lines, and greater than the instrumental width (Sect.\, \ref{DR}). In addition, the ratio between the equivalent widths of the two lines, $EW_{\rm \lambda5890}$/$EW_{\rm \lambda5896}$, is restricted to vary from  2 (i.e., optically thin absorbing gas) to 1  (i.e., optical thick limit)  according to \cite{Spitzer1978}. 
 In general, the two-component modeling of the NaD line profile is rather complicated and often leads to unphysical or non-unique solutions. Therefore, in order to preserve against spurious results, the secondary (stellar) component was constrained to be within $\pm$\,100 \kms \ of the systemic rotating velocity as traced by the \co \ and \Ha \ features\footnote{As discussed in Sect.\, \ref{bulk}, the bulk of the ionized gas and the stars nearly co-rotate. Therefore we linked the secondary NaD component  to \Ha \, rather than to  \co, since those features were obtained from the same data-cubes and therefore have homogeneous spatial sampling and calibrations.}.
We checked the line profile fits by visually inspecting the residuals. A fit with two-kinematic components was preferred when it produced a significant reduction of the residuals with respect to the standard one-component fit. In a few spectra we also modeled the HeI $\lambda$5876 nebular emission line (15 $\AA$ blueward of NaD) since it was broad and strong enough to affect the final results of the NaD fitting. For each  line and component of the doublet, the output of the fitting process gives the  central wavelength, the line width, and the absorbed flux along with their respective fitting errors. The intrinsic line widths were computed after removing the instrumental profile inferred from the sky lines (see Sect.\, \ref{DR}). The error on the relative central wavelength was  obtained combining in quadrature those associated with the fitting and the calibration, giving a total error of $\Delta\epsilon$\,=\,$\sqrt{\Delta\lambda_{\rm cal}^{2}\,+\,\Delta\lambda_{\rm fit}^{2}}$. For high S/N spectra the calibration errors (typically on the order of $\sim$\,0.28\,$\AA$) are dominant, while in the low S/N regime the fitting errors  (generally smaller than 0.32 $\AA$) are more important.   Figure \ref{spaxels}  shows examples of the one- and  two-Gaussian fits.  \\

\label{NaDLF} 

%  Two kinematics components were preferred when the fitting routine produce a significant reduction  (i.e., $>$\,10\,$\%$) of the $\chi^2$}.

 \subsubsection{H$\alpha$ emission line profiles}
 
\begin{figure}%[b]
		\centering
		\includegraphics[scale=.64]{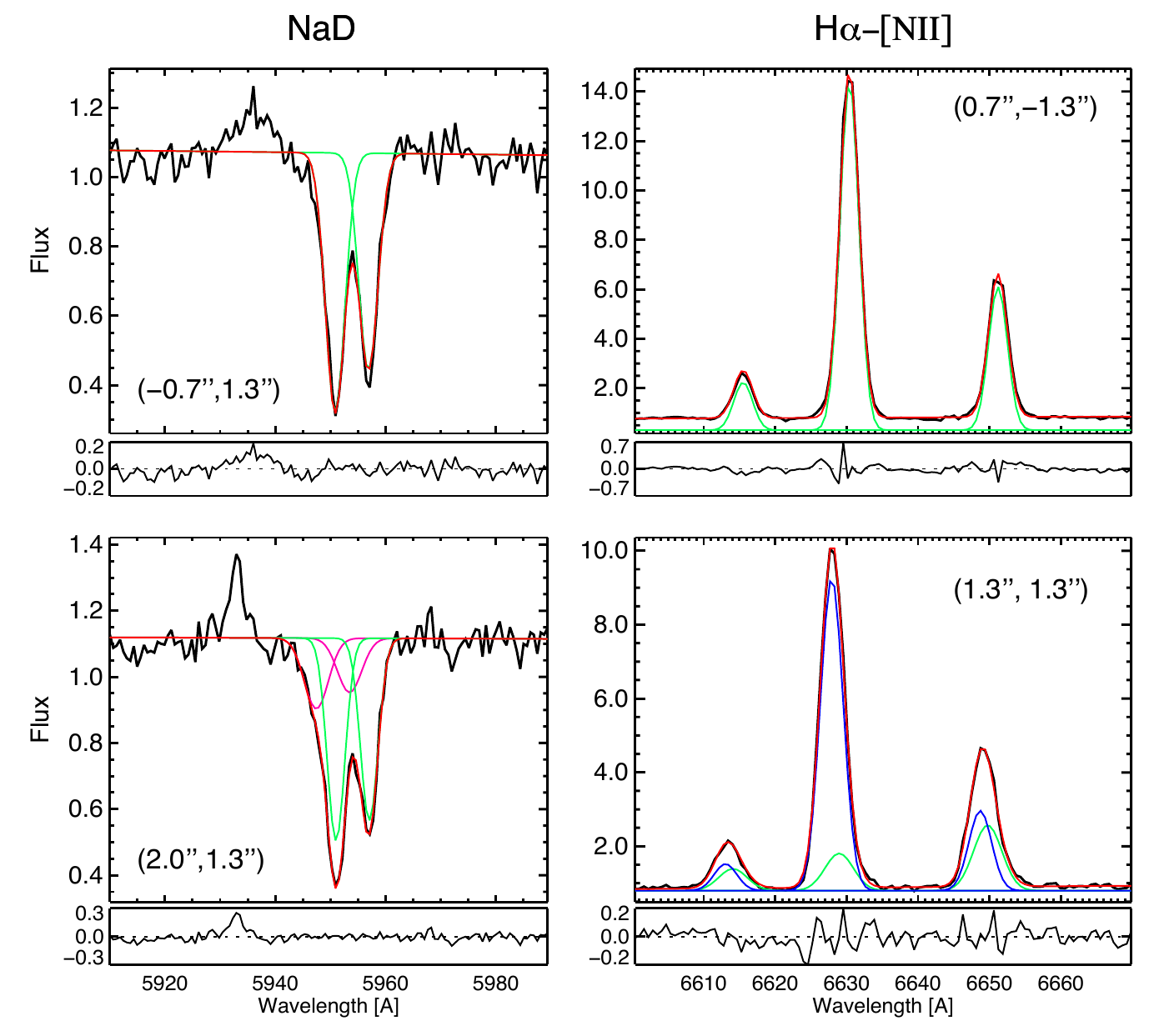}
 		\caption {NaD  (left) and   \Ha \,-[NII]  (right) spectra for  selected regions. In the  labels the coordinates have been indicated as the distance from the nucleus (map center, Figures \ref{NaDpanels} and \ref{Hapanels}).  For each spaxel the modeled  line profile  (red line) and  the single component (with different colors) are  shown, along with the residuals (small panels). \textit{Top:} The results of the NaD  and the \Ha \, one-component modeling.  The green lines in each panel represent, respectively, the main-NaD and the narrow-\Ha \, Gaussian functions. \textit{Bottom:}  Examples of NaD and \Ha \, spectra with two components. The secondary-NaD and the broad-\Ha \, curves are shown in magenta and blue,  respectively, and  the  green and red lines represent the main component and the global fits to the observed profiles (as in the top panels). }	
\label{spaxels}
	\end{figure}

For the H$\alpha$-[NII] emission line complex we use the same approach as for the NaD. Therefore, we initially fit the spectra with a single Gaussian per line analyzing the residuals and look for the possible existence of a secondary component. This method is similar to that carried out by \cite{Bellocchi2013}, although here a more detailed spaxel-by-spaxel search for a secondary component has been followed (something unpractical for the large sample analyzed by Bellocchi et al.). \\
Most of the spectra were accurately fitted by a single Gaussian per line although a small fraction of spectra, mainly in the inner region (and towards the NE), show traces of a secondary component. The IDL-fitting routine was customized to model the \Ha \, and [NII]$\lambda\lambda$6548, 6583 $\AA$ emission lines, and to estimate the flux intensity, central wavelength, and widths. Specifically, the  \Ha-[NII] complex was fitted with one (or two)  Gaussian(s) per line with the line flux ratios and wavelengths of the [NII]-lines  fixed according to the atomic physics. Similarly to the NaD doublet, the widths are constrained to have a velocity dispersion greater than the instrumental resolution (Sect.\, \ref{DR}) and to be equal for all the lines of the same component. Typically, the  \Ha-fitting errors are lower than 0.1 and 0.15 $\AA$ for the central wavelength and the width respectively, while the calibration errors are the same as those adopted for the NaD.  Figure \ref{spaxels} shows examples of one- and two-Gaussian fits, respectively. The two-Gaussian fits  lead to two components which in general can be distinguished according to their widths, i.e.,  narrow ($\sigma$\,=\,46\,$\pm$\,10 \kms) and broad ($\sigma$\,=\,74\,$\pm$\,5 \kms) components. The narrow (main) component contains most of the \Ha \, flux and it is associated with systemic rotational motions while, the broad component seems to be associated with non-rotational motions (see Sect. \ref{kingeom}).

\label{HaLF} 
	
  \begin{figure}%[h]
		   \centering
	 	   \includegraphics[scale=0.43]{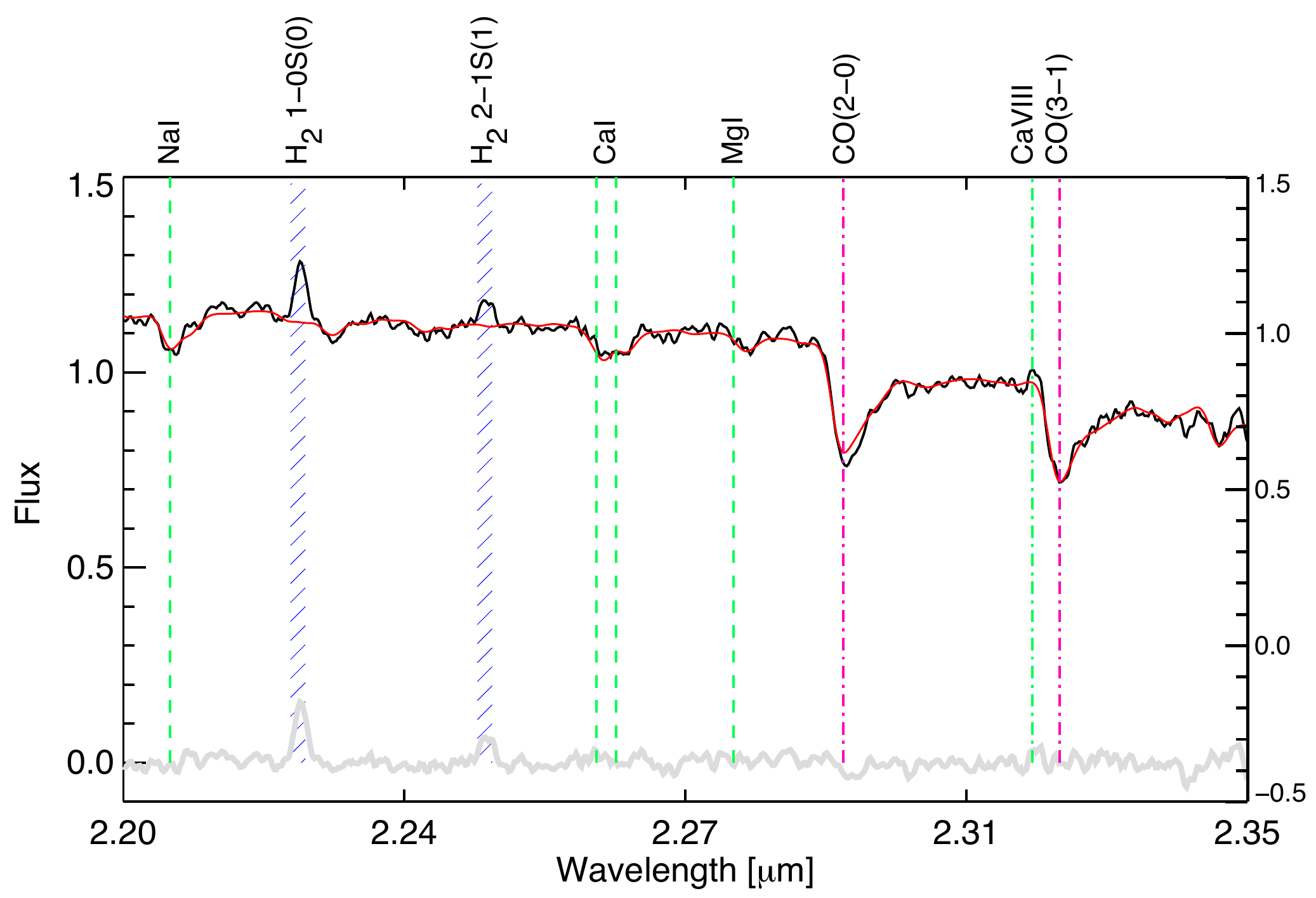}
		    \caption{The near-IR spectrum (black) from the central bin (with S/N\,$\sim$\,27) of \obj \, with overlaid its best-fitting template spectrum derived with  the pPXF approach (Sect.\, \ref{otherLF})  is shown (red line; fit residuals are shown in gray). The location of the wavelength bands masked  during the fitting  are marked in dark blue. The most relevant spectral features are labeled at the top and marked with a dashed green line, while the CO bands are marked with a magenta dotted-point line. The residuals from this process are low, typically within -0.1 and 0.2.}
		    		       \label{COspaxel}		
		\end{figure}
		
 \subsection{Near-IR stellar absorption features}

We also obtain the stellar kinematics traced by the near-IR CO(2-0)$\lambda$2.293\,$\mu$m and CO(3-1)$\lambda$2.322\,$\mu$m absorption bands, using a maximum likelihood approach, i.e., the \textit{P}enalized \textit{P}i\textit{X}el-\textit{F}itting method (\textit{pPXF}; \cite{Cappellari04}), to fit a library of stellar templates to our SINFONI data (Fig.\,\ref{COspaxel}). In particular,  we use the library from \cite{Winge2009}  whose spectra cover the wavelength range 2.15\,--\,2.43$\mu$m with a spectral resolution and sampling of R\,$\sim5600$ and 1\,\AA\,pixel$^{-1}$, respectively. These spectra are convolved to match the SINFONI resolution before the fitting procedure is applied. We find typical errors for velocity and velocity dispersion, calculated as the 1-$\sigma$ error, of less than 15 \kms. Before extracting the kinematics, we perform an adaptive spatial binning using the Voronoi method by \cite{Cappellari04}. As outlined in Piqueras L\'opez et al. (2012), with this technique we maximize the S/N over the entire FoV using bins of approximately circular shape in low S/N regions, preserving the spatial resolution of those regions above a minimum S/N threshold (i.e., 25).

 \label{otherLF}

 \subsection{2D Maps: Summary of components}
 
\begin{figure*}
			\centering
	 	 	  \includegraphics[scale=0.76]{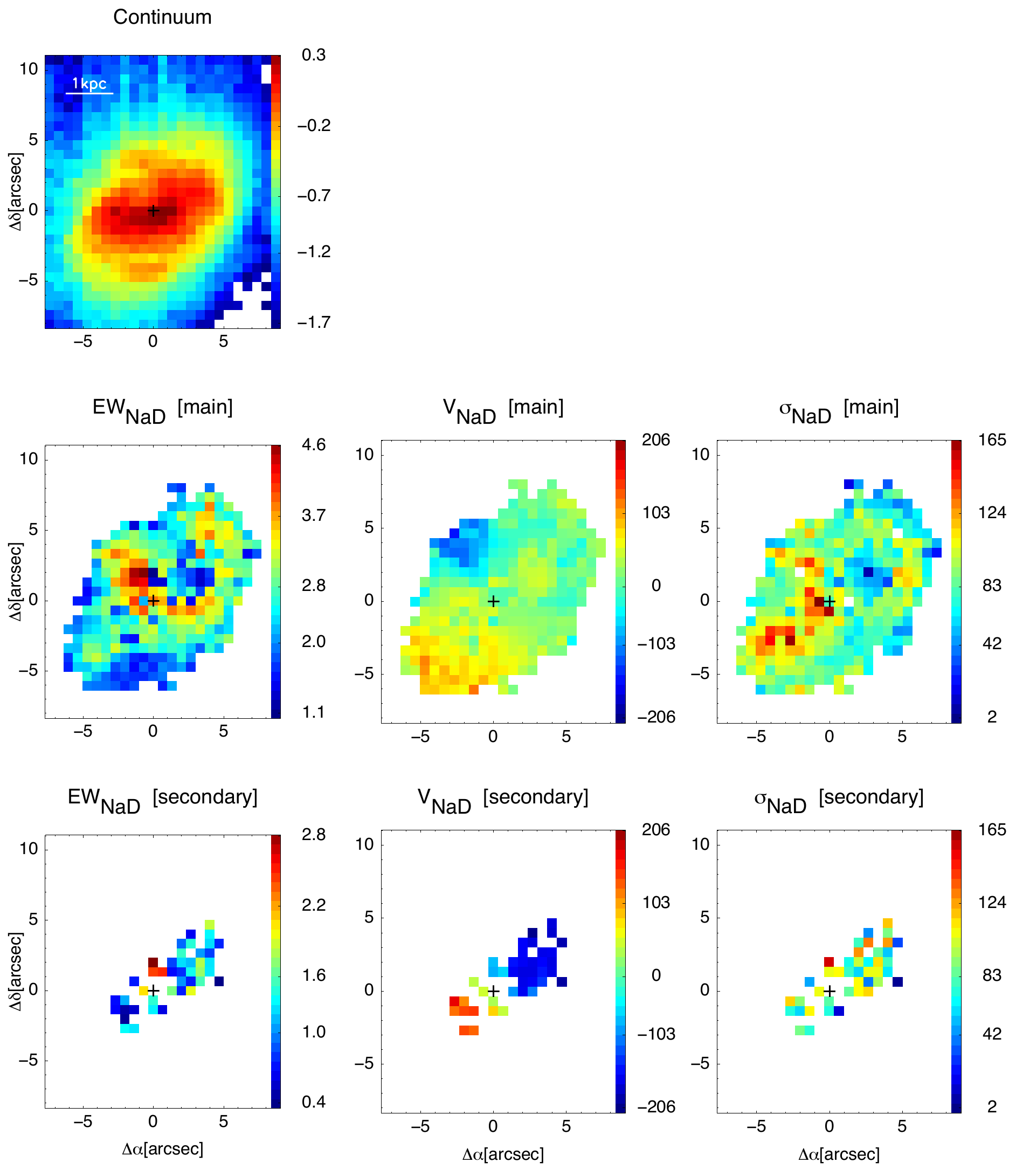}%{morphlett.pdf}%{mappe/vel_grandi/morph.pdf}
			  \caption{Neutral  gas maps as derived from the NaD absorption line profiles.  \textit{First row:} optical continuum image (represented in logarithmic scale and in units of 10$^{-13}$erg s$^{-1}$ cm$^{-2}$) obtained from  the mean of the line-free continuum nearby the doublet  in a 100 $\AA$\, restframe-wavelength range. \textit{Second row:} the equivalent width map  for the $\lambda$5890 line,  velocity field, and velocity dispersion map   for the main component. \textit{Third row:} the same as the second row but for the secondary component. EWs maps are in  $\AA$, while the kinematic maps are in \kms. %Note that, these maps have been zoomed-in since the VIMOS FoV  is wider  (by a factor $\sim$ 2) than the area covered by the NaD absorption.   
			  The cross marks the galaxy nucleus (Sect.\, \ref{maps}).}	
			\label{NaDpanels}
		\end{figure*}

\begin{figure*}
			\centering
	 	 	  \includegraphics[scale=0.78]{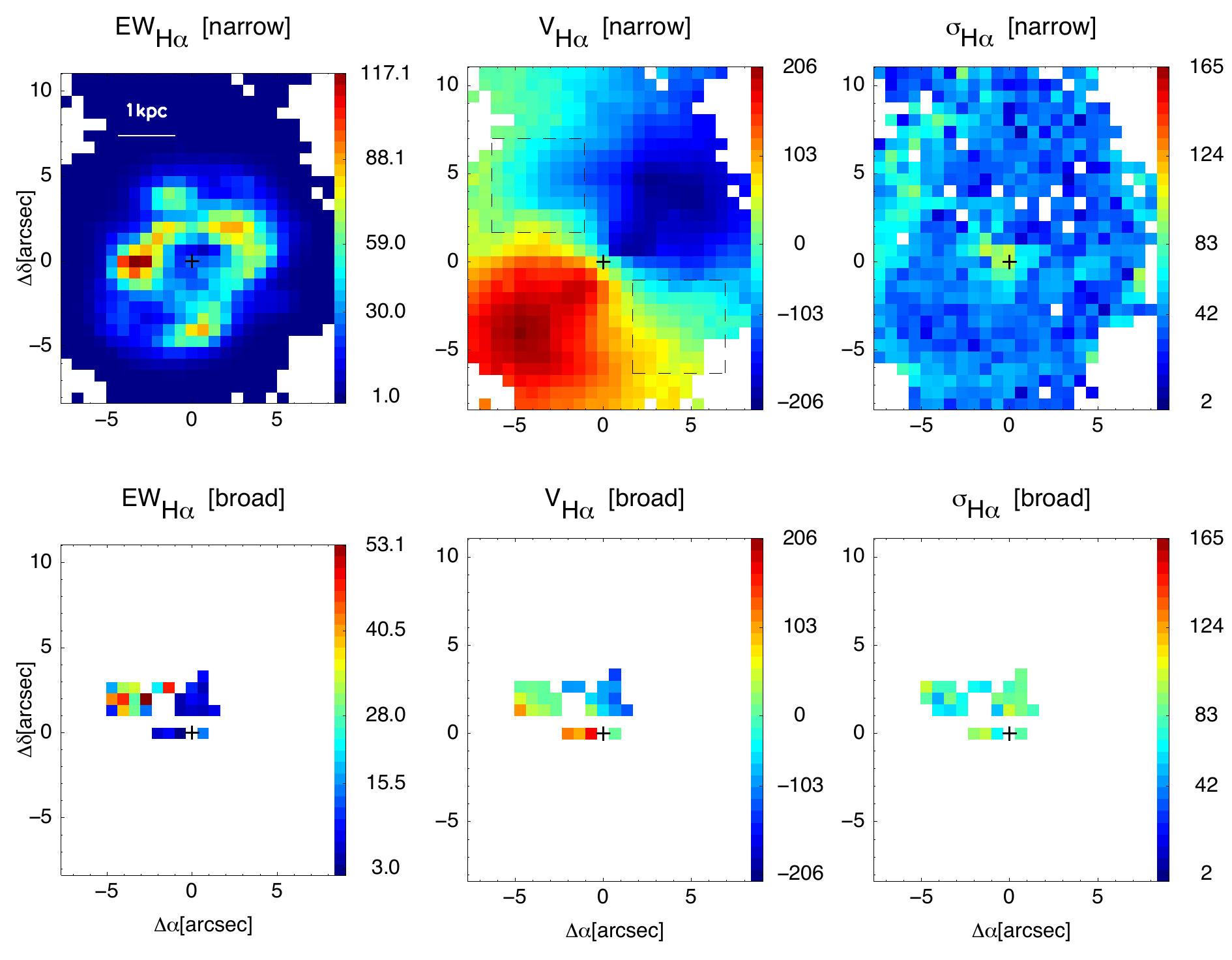}%{morphlett.pdf}%{mappe/vel_grandi/morph.pdf}
			  \caption{Ionized  gas maps as derived from the \Ha \, emission line profiles.  \textit{First row:} the equivalent width map, velocity field, and velocity dispersion map  of the main (narrow) \Ha \, component. \textit{Second row:} the same as the first row but for the secondary (broad) component. EWs maps are in $\AA$, while the kinematic maps are in \kms.  We note that these  maps have been zoomed-in for consistency with the NaD maps.  The cross marks the galaxy  nucleus, as in Fig.\,\ref{NaDpanels}. The black dashed boxes in the velocity map (narrow component) mark the regions where the integrated spectra are obtained and analyzed (see Sect.\,\ref{kingeom}).}
			  \label{Hapanels}
		\end{figure*}

\begin{figure*}%[b]
		\centering
         	\includegraphics[scale=.95]{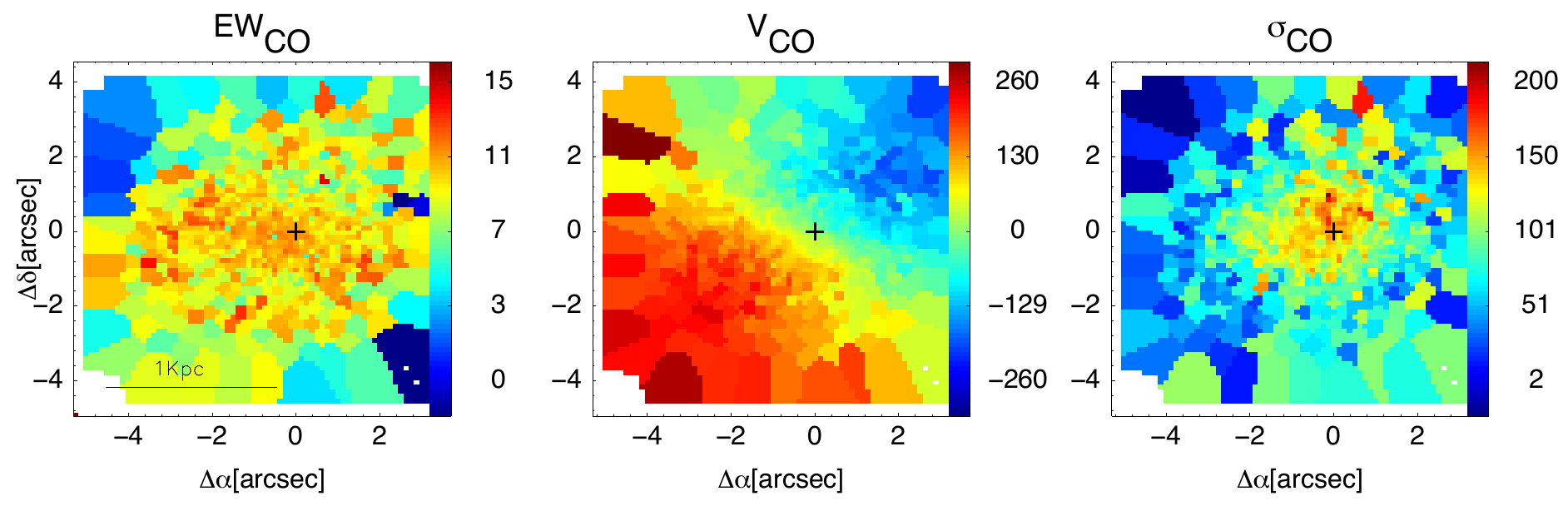}
 		\caption{From left to right: equivalent width given in $\AA$, velocity field   and velocity dispersion  \textit{(c)} maps, represented  in units of \kms, of the stellar component in  \obj \, as traced by the \co \, $\lambda$2.293\,$\mu$m infrared absorption line (\Piquerasprep).   The  cross  marks the  nucleus (see Sect.\, \ref{maps}). As mentioned in Sect.\, \ref{otherLF},  data were binned using the adaptive Voronoi  2D-binning method, developed by \cite{Cappellari04}. We note that the ring-like structure with height \Ha-EWs (Fig.\,\ref{Hapanels}) is absent in the EW(CO) map.}
		\label{COmaps}
		\end{figure*}

Figures \ref{NaDpanels}, \ref{Hapanels}, and \ref{COmaps}  show the spatial distribution of the EWs, the velocity fields, and the velocity dispersion maps of the different kinematic components obtained from the NaD doublet, the \Ha \  line, and the CO bands, respectively. In Fig.\,\ref{NaDpanels} we also include for reference the continuum (5910-6110$\AA$) image derived from the VIMOS data cubes. The NaD and \Ha \ velocity maps consider as zero-velocity inferred from  \Ha \ at the optical nucleus (i.e., 3110 $\pm$ 5 \kms). We assume that the optical and the near-IR nuclei (i.e., intensity peak) are in positional agreement within the spatial resolutions of the two sets of data. All the maps are displayed using the plotting package \textit{jmaplot} (\cite{Maiz04}).  \\
In Table \ref{Velocities},  we summarize  some kinematic parameters of the components traced by the different spectral features used in the present paper (NaD,  \Ha, and the \co). It is important to note that these spectral features trace the properties of different constituents, and their components do not necessarily correspond to each other.  As we will discuss through the paper,  the  NaD absorption has three contributors at different locations along the line of sight. Specifically, the main NaD component consists of neutral ISM clouds likely forming an irregular disk-like structure and a galactic wind emerging perpendicular to this disk (see Sect.\, \ref{kingeom}),  while the secondary NaD component is likely associated with the stars. In contrast, the  main component of \Ha\ emission probes the ionized gas systemic rotational motions, and the secondary (broad) component seems associated with radial motions. The \co \ univocally traces the systemic rotational velocity of the stellar component.

 \label{maps}
 %It is represented  in erg s$^{-1}$ cm$^{-2}$  $\AA^{-1}$ flux units  \textbf{Check} and

\begin{table*}[t]	
					\caption{Main kinematic properties of the different constituents/phases of \obj\ }				
				\begin{center}
				%\resizebox {\textwidth}{!}{%
				{\begin{tabular}{ l  l l c c c c c }
					\hline
				\hline
				Constituent/Phase & Tracer & Component (structure) &  &   $\Delta V^{a}$& $\sigma_{\rm c}^{b}$ &$\sigma^{c}$ & PA$^{d}$  \\
				                                  &               &                       &  &   \kms \,               &     \kms \,        &        \kms \,                                        & degree \\
				\hline
				
				  Stars                        & \co        & main (disk) &   & 188 $\pm$ 11 & 136 $\pm$ 20 & 78 $\pm$ 16 &139\\
				                                   &   NaD   &  secondary (disk)  & &  150 $\pm$ 15 & 152 $\pm$ 31 &  82 $\pm$ 31 &127 \\
				  \hline
\multirow{2}{*}{Ionized Gas} & \multirow{2}{*}{\Ha}  & main / narrow (disk) &  & 203 (202) $\pm$ 4 & 95 $\pm$ 4& 45 (44) $\pm$ 10  & 135  \\
                                                       &                                  & secondary /broad  &  & 145 $\pm$ 12 &  ... &74 $\pm$ 5 & 60\\
                                             \hline
\multirow{2}{*}{Neutral Gas} & \multirow{2}{*}{NaD} & main (disk) &     & 83 (80)$\pm$ 12 & 127 $\pm$ 5&  84 (83) $\pm$ 10& 129 \\
                                                   &                                       & main (outflow) &  &  -64 $\pm$ 20 & ...  & 98 $\pm$ 10 & 45\\

				\hline
					\hline
				\end{tabular}%}
				\label{Velocities}}
						\end{center} 
				\tiny{Notes. --- $^a$  observed velocity amplitude; $^b$  nuclear velocity dispersion ($\sigma_{\rm c}$),  $^c$ mean  velocity dispersion ($\sigma$), and $^d$ position angle of the major kinematic axis.  For the main-NaD and -\Ha\  components, the values in brackets are derived over the same FoV as those for the CO  (i.e., within a radius of R $\leq$ 1.1 kpc).}			 	
			\end{table*}

  \section{Results and discussion}
	\label{res}

	   \subsection{Stellar and ionized gas kinematics and dynamic structure}	

\begin{figure}[h]
		\centering
        	\includegraphics[scale=.52]{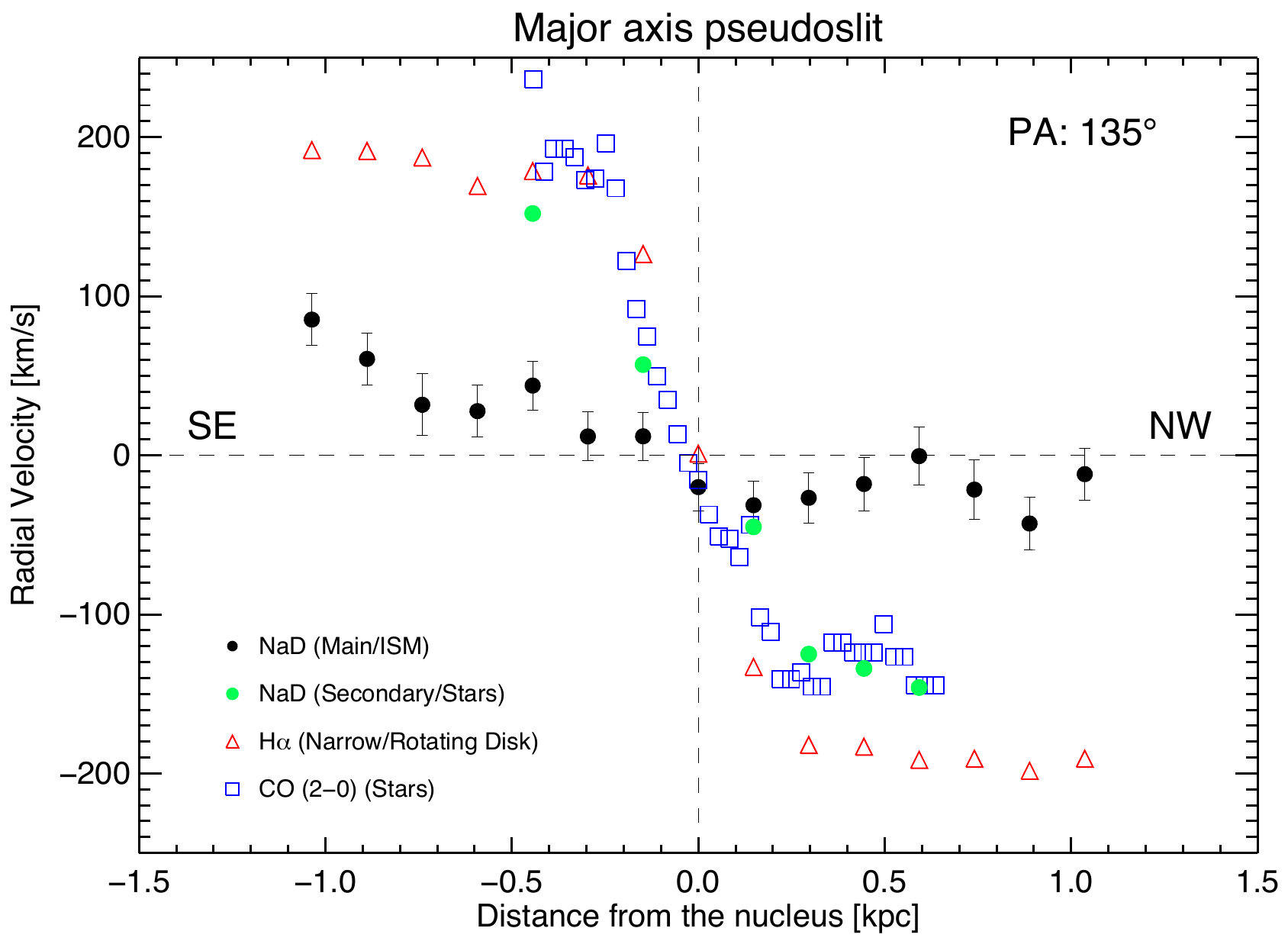}
 		\caption{PV diagram along the major axis (PA\,$\sim$\,135 degrees)  for different tracers including NaD  (the black and green circles indicate the main and the secondary components, respectively, see  Sect.\, \ref{NaDLF}),  narrow \Ha \, (red triangles), and \co \,  (blue squares) data. Error bars are not displayed for the ionized gas, stars, and the  NaD secondary component speeds since they are, typically, less than 15 \kms, therefore smaller than the corresponding symbols. We note that we did not include the PV curve for the \Ha-broad component since it is not observed  along the major kinematic axis and  it is likely tracing non-circular motions  (Sect.\, \ref{kingeom}).}
		\label{DVP}
		\end{figure}

The velocity fields of the ionized gas and stars, as  traced by  the \Ha \ (narrow component) and the \co \ line (Figures \ref{Hapanels} and \ref{COmaps}, respectively), present point-antisymmetric velocity patterns consistent with large, kiloparsec-scale ordered rotational motions (e.g., rotating disks). The ionized gas major kinematic axis is well aligned with that of the stars, and both are in turn coincident within uncertainties with the photometric axis as seen in the \textit{HST} images (Fig.\,\ref{immagine}), with PA\,$\sim$\,135$\degr$ $\pm$  5$\degr$. The  position-velocity diagram (PV diagram, Fig.\,\ref{DVP}) highlights that the stars and the ionized gas have similar rotation, although the latter has a slightly larger amplitude\footnote{The velocity amplitudes are obtained as half the peak-to-peak velocity difference within a radius of $\sim$\,1.1 kpc on either side of the galaxy.} (i.e., $\Delta$V(H$\alpha$)\,=\,203\,$\pm$\,4 \kms \, vs. $\Delta$V(CO)\,=\,188\,$\pm$\,11 \kms). It is interesting to mention that the observed star forming circumnuclear ring structure (R $\sim$ 0.7 kpc) identified by the large  \Ha-EWs (Fig.\,\ref{Hapanels}) follows the global ordered motions since no kinematic asymmetries at the ring's position are observed in the \Ha \ velocity map.\\
As expected for a rotating disk, the \Ha-velocity dispersion map shows a centrally peaked pattern with its maximum value (i.e., $\sigma_{\rm c}$(\Ha)\,=\,95$\pm$\,4 \kms) in positional agreement within uncertainties with the nucleus. The extra-nuclear (R $\geq$\,0.3 kpc) mean velocity dispersion (i.e., 45  $\pm$ 10 \kms) is significantly larger than that found in less active local star forming spirals (e.g., GHASP survey, $\sigma$\,$\sim$\,24 \kms; \cite{Epinat2010}). Although this value is not as large as those found at high-z ($\sigma$\,$\sim$\,60-90 \kms; \cite{ForsterS2011}), it indicates that the ionized disk is thicker than in normal spirals. Thick disks seem common in [U]LIRGs as discussed in \cite{Bellocchi2013} and Arribas et al. in prep). In addition to a relatively large mean value, the ionized gas $\sigma$-map also reveals the presence of some large-scale extranuclear regions towards the NE (R $\geq$ 1 kpc,  PA from 0$\degr$ to -110$\degr$) with values clearly above the average ($\sigma$\,$\sim$\,50 -100 \kms). These can be explained by an increased turbulence or by the presence of an extra non-rotational component in these regions (see Sect.\, \ref{kingeom} for details). \\
Similarly to the ionized gas, the stars exhibit a centrally peaked velocity dispersion map (Fig.\,\ref{COmaps}c), but  the nuclear  (R\,$<$\,0.3 kpc) velocity dispersion is slightly larger than that of the ionized gas (136 $\pm$ 20  vs.  95 $\pm$ 4 \kms). On the one hand, these findings support the idea that the stellar and ionized gas central velocity dispersions are dominated by the gravitational potential of the galaxy (i.e., its mass). On the other hand, the substantially large stellar extra-nuclear, within 0.3\,$<$\,R\,$<$\,2 kpc, mean velocity dispersion value  (i.e., 78 $\pm$ 16 \kms), indicates an extra-dynamical support for stars with respect to the already turbulent ionized gas disk. The low rotational  support of the stellar disk with respect to that of the ionized gas can also be drawn from the observed (i.e., no inclination corrected) velocity-to-velocity dispersion (V/$\sigma$) ratio, calculated as the ratio between the amplitude and the mean velocity dispersion across the disk.  Specifically, for stars this ratio (i.e., 2.4) indicates a larger dispersion component while that of the ionized gas (i.e., 4.5) is evidence of a rotation-dominated kinematics.\\
Assuming, as in Cresci et al. (2009), that the stars and the ionized gas are distributed in  thin rotating disks, the scale height of the disks ($h_{\rm z}$) can be derived as  (\cite{binney2008})
\begin{center}
	 \begin{equation} 
h_{\rm z}= \frac{\sigma^{2}\times R}{V(R)^{2}}.
	\label{thindisk}
	\end{equation}
  \end{center}
  
Considering the projected semi-amplitude (V(R)) and the mean velocity dispersion\footnote{This choice implicitly assumes that the velocity dispersion  is mainly due to the gravitational potential  of the galaxy rather than to turbulence. Therefore, these estimates of heights  represent upper limits.} across the galaxy disk (i.e., $\sigma$)  for the stars and the ionized gas, heights of 125 and 35 parsecs are obtained. These heights were obtained at a distance \textit{R}\,=\, 2 kpc  from the nucleus, which is the radius at which the ionized gas and the stellar rotation curves flatten (as seen in Fig.\,\ref{DVP}). \\
In summary, the stars and the ionized gas nearly co-rotate. Their kinematics are consistent with the picture of a (mainly) rotationally supported ionized gas disk of about 35 pc (half thickness) embedded in a thicker and dynamically hotter stellar disk of about 125 pc. The ionized gas shows a relatively high turbulence compared with normal spirals of lower SFR, but it is not as high as that found at high-z. 
%It is confined in a mainly rotationally supported disk embedded in a thicker and dynamically hotter stellar disk with a non-negligible random-motion component. 
\label{bulk}

	%\textit{Considering that the scale height, h$_{\rm z}$, is proportional to  the observed  $\sigma$ /$\Delta$V ratio under the thick disk approach (e.g., \cite{Cresci2009}), the stellar disk should be about a factor 1.5 - 2 thicker  than that for the ionized gas.}\textbf{We should remove this sentence and we should insert the other} \\ 	The lag in velocity and the larger velocity dispersion of stars  with respect to that of ionized gas implies a larger scale height and lower rotational  support of the stellar disk than that of the ionized gas. 
	%
%dynamical status of the ISM is more turbulent, and the ionized disk is thicker than in normal spirals.  Turbulent and thick disks seem common in [U]LIRGs as discussed in \cite{Bellocchi2013} and Arribas et al. in prep).
	
		%we infer that the stellar disk is thicker than that of the ionized gas: 35 vs. 125 pc.

		\subsection{The interstellar origin of most of NaD absorption: disentangling the stellar contribution}

\begin{figure*}
		   \centering
	 	   \includegraphics[scale=0.6]{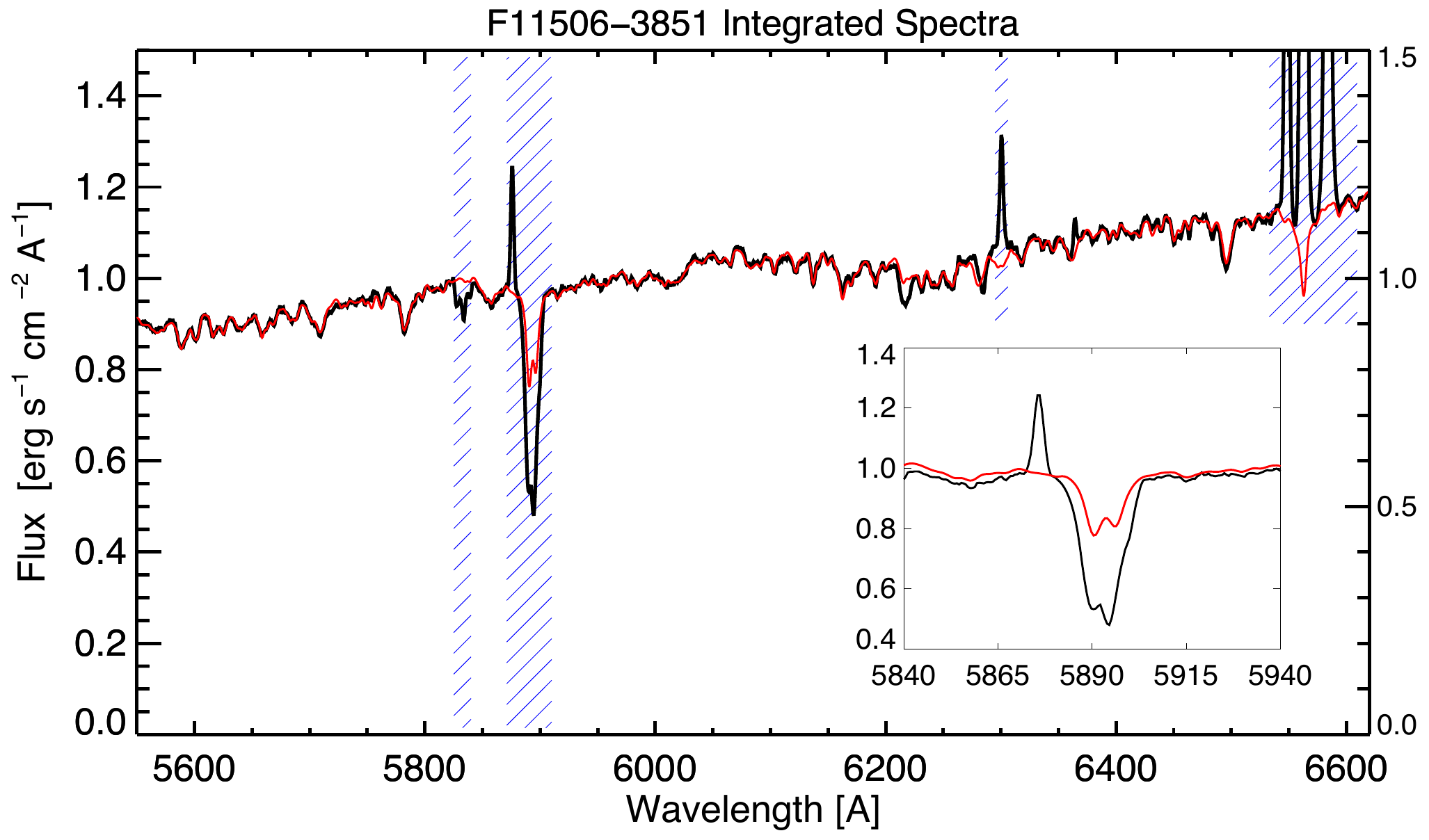}
		    \caption{The integrated spectra (black) and  its best-fit stellar spectra (red) obtained with  the pPXF approach (Sect.\, \ref{stellarNaD})  are shown. The gas emission wavelength bands masked  in the fitting  are marked in dark blue. The  insert highlights  the NaD region with a zoomed view. The lack of good agreement in the doublet feature indicates that the NaD  extra-absorption is interstellar in origin.}    	
		    \label{ppxf}
		\end{figure*}

The NaD doublet  originate in clouds of warm neutral gas (i.e., interstellar absorption) and in late-type stars (e.g., K-type giants, \cite{Jacoby84}). Therefore, as a first step towards interpreting this feature, we need to determine the respective roles of these two potential contributors (i.e., stars, and neutral gas). One method for evaluating the global stellar-NaD contribution consists in fitting a stellar spectra to our data. As this approach requires a relatively high S/N, we first generate  spatially integrated spectra  following the S/N optimization method described in Rosales-Ortega et al. (2012). We then use the pPXF method (\cite{Cappellari04}) with the Indo-U.S. stellar library (\cite{Valdes04}) to produce a model spectrum that matches the observed stellar continuum, after masking known gas emission lines and a $\sim$\,30\,$\AA$\, region centered in the NaD feature. The result of this approach is a model that in general reproduces accurately the  continuum shape (see Figure \ref{ppxf}). However, the (purely stellar) modeled NaD profile leaves a strong residual when compared with the data (insert in Fig.\,\ref{ppxf}). Specifically, the modeled stellar-NaD spectra has $EW_{\rm 5890}$\,=\,(1.1\,$\pm$\,0.2) $\AA$, which correspond to only a small fraction ($\sim$\,1/3) of the total observed absorption,  suggesting that it is mainly ($\sim$\,2/3) interstellar in origin\footnote{This determination is in excellent agreement with the EW(NaD)  derived from the relation between the stellar MgIb$\lambda$5174 and NaD  (i.e.,  EW(NaD)\,$\sim$\,0.75 EW(MgIb), \cite{H00}), considering the median value of EW(MgIb) for the HII-selected luminous \textit{IRAS}-galaxies sample by \cite{KVS} (i.e.,  1.1 $\AA$).}.\\
The stellar continuum fitting method described above cannot be applied on a spaxel-by-spaxel basis, as the individual spectra lack the required S/N. However, the spectral fitting decomposition described in Sect.\,\ref{NaDLF} suggests that the main NaD component is due to neutral gas absorption, while the secondary component is associated with the stars. On the one hand, the EW-map for the main component (Fig.\,\ref{NaDpanels}) has relatively large values  (typically $\geq$\,1.5 $\AA$) and it exhibits a patchy and off-centered morphology that contrasts with the one for the continuum  (Fig.\,\ref{NaDpanels}). In addition, the slow rotating velocity field and the off-centered velocity dispersion map associated with this component significantly differ from those for the stars (see Fig.\,\ref{COmaps}). On the other hand, the secondary component of NaD has a velocity field (Fig.\,\ref{NaDpanels}) that closely follows that of the stars (Fig.\,\ref{COmaps}),  its nuclear and mean line-width are in good agreement with those for the \co \, (see Table 2), and its mean EW (i.e.,  1.2\,$\pm$\,0.5 \,$\AA$) is consistent with a stellar origin. \\ 
In summary, the application of the stellar-continuum modeling approach to the integrated spectrum indicates that most ($\sim$\,2/3) of the NaD absorption in \obj \ is interstellar in origin, although there is a significant contribution ($\sim$\,1/3) due to the stars. This is also supported by the spaxel-by-spaxel spectral fitting decomposition, according to which we have been able to disentangle the 2D structural and kinematic properties of two kinematically distinct components. While one of these components dominates the absorption and  is likely tracing the neutral gas in the ISM,  the secondary weak component has properties fully consistent with a stellar origin. 

 \label{stellarNaD}

 \subsection{Morphology, kinematics and dynamical support of the neutral ISM}
 
 As discussed in the previous section, the neutral gas is traced by the main component of the NaD doublet. Its EW-map (Fig.\,\ref{NaDpanels}) reveals relatively strong absorption (i.e.,  NaD-EWs  of 2.5-4.5 $\AA$) at projected distances of R\,$\sim$\,0.6 - 1.2 kpc from the nucleus and it has an irregular and patchy morphology, which is unrelated to that of the optical continuum (Fig.\,\ref{NaDpanels}).\\
The overall neutral gas velocity field (Fig.\,\ref{NaDpanels}) lacks the typical pattern of a rotation dominated system. The deviations from an ideal rotating disk include poorly defined kinematic axes,  a small velocity amplitude  (i.e.,  83\,$\pm$\,12 \kms), and an asymmetric global structure with respect to the galaxy nucleus\footnote{Part of the irregularities observed in NaD velocity map are a consequence of its relatively small amplitude, which magnifies the effects of the errors in the overall structure of the map.}. Moreover, contrary to what it is expected for a rotation-dominated gas disk, velocities strongly blue-shifted with respect to the systemic (by up to -154 \kms) are observed along the minor axis towards the NE. While in Sect.\, \ref{kingeom} we give an explanation for the kinematic properties found along the semi-minor axis, in this section we will focus on the general properties of the kinematic maps of the main NaD component excluding that particular region. \\
Therefore, leaving aside the region with large blueshifted velocities, the velocity field of the main component of NaD (i.e.,  neutral gas) has a pattern consistent with a slowly rotating (irregular) disk.  Its comparison with the H$\alpha$ velocity field (Fig.\,\ref{Hapanels}) indicates that the neutral disk lags in velocity compared to the ionized gas. The neutral and warm gas components have significantly different amplitudes (i.e.,  83 vs.  188 \kms, respectively; Fig.\,\ref{DVP}), indicating that they are not kinematically coupled.  Part of the above mentioned velocity-discrepancies may be due to the different location of these ISM phases. While the H$\alpha$ emission is associated with dense gas regions, the neutral gas NaD clouds could be located arbitrarily between the observer and the background light (the stellar disk) along the line of sight. A different neutral gas 3D spatial distribution with respect to that of the ionized ISM and stars can be drawn from the comparison of their velocity dispersion maps. In Fig.\,\ref{NaDpanels}, it is clear that the neutral gas dispersion values  are larger than those measured for the ionized gas (typically by more than 50  \kms \, excluding the innermost region) indicating that the neutral gas is located in a thicker disk. The different pattern between the velocity dispersion maps for the stars and the neutral gas also suggests that they have different spatial distributions.\\
We derive the scale height ($h_{\rm z}$) of the neutral disk  adopting the thick disk model outlined in \cite{Cresci2009}.  This differs from the thin disk (Eq.\,\ref{thindisk}) as the $h_{\rm z}$ scales with the  mean velocity dispersion across the disk ($\sigma$) and the rotational velocity ($\Delta V(R)$, i.e.,  the de projected semi-amplitude) at radius \textit{R} (i.e.,  2 kpc as in Sect.\ref{bulk}): $h_{\rm z}$\,=\,$\sigma$\,$\times$\,$R$\,/\,V (\cite{genzel08}). In this case, the height scale for the neutral gas is $h_{\rm z}$\,$\sim$\,700 pc, confirming that the neutral gas is distributed in a thick disk. However, this estimate represents a lower limit (by a factor of 2 or more) given the potential contribution of unresolved random motion that could be supporting part of the mass and thickening the disk. We note that the ISM absorption features studied here  only probe the gas clouds in front of the stars. Therefore, even if a small scale-height neutral disk is present, it would not be detected because many of the stars,  contributing to the continuum light,  are in front of it.\\
In \obj, the comparison of the  V/$\sigma$ ratios  for the different ISM phases and the stars indicates different levels of dynamical support. The neutral gas has a dispersion-dominated kinematics (V/$\sigma$\,$\sim$\,1) while the stars and the ionized gas ($V/\sigma$\,=\,2.4 and 4.5, respectively) have an increasing rotational component  (e.g.,  disks). The low rotational support for the neutral gas is compensated by the large random motions observed,  and it implies that the neutral gas is distributed in a thick disk as observed. \\
Summarizing, the  NaD-EW map suggests a rather irregular and complex distribution of the neutral gas. Excluding a region of strong blueshifted velocities along the semi-minor axis, the overall neutral gas kinematics can be interpreted as a slow rotating disk that lags in velocity to those of the ionized gas and stars,  and it is dominated by random motions as indicated by the small V/$\sigma$ value. Therefore, neutral gas is distributed in a thick and dynamically hot disk, in contrast to the stars and the ionized gas, which are confined to disks with lower height scales mainly supported  by rotation. 

\label{prop}

%\footnote{\textbf{A similar results has been obtained for the nearby barred spiral NGC 891. \cite{Fraternali2005} show the presence of extra-planar neutral HI-gas up to distances of 15 kpc from the galaxy plane.}}

\subsection{The wind kinematics and geometry}

     \begin{figure*}
		   \centering
		   \includegraphics[scale=0.9]{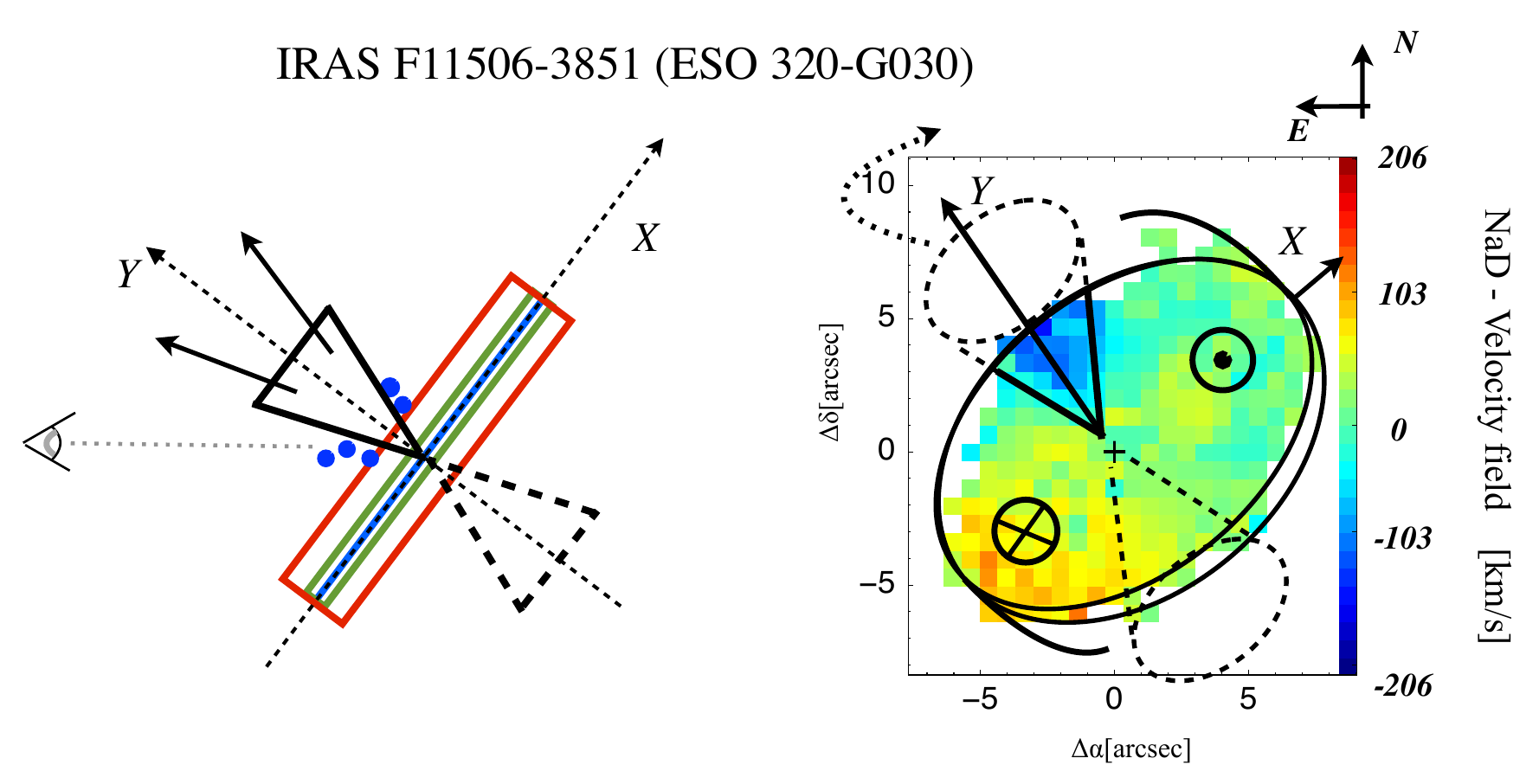}
		    \caption{Geometric model of a conical wind perpendicular to the galaxy disk.   In both images, the black lines represent the wind cones. The receding cone (invisible  for the observer in NaD,  because of the lack of background light, i.e.,  the stellar disk) is displayed with dashed lines. The  letters \textit{X} and \textit{Y} mark the major and minor axes of the galaxy.  \textit{Left:} the red, green, and blue boxes represent, respectively, the neutral, stellar, and ionized disks (see Sects.\, \ref{bulk} and \ref{prop}). Blue filled circles indicate the wind ionized phase (see Sect.\,\ref{kingeom}  for details).   \textit{Right:} sketch of the bipolar GW and the spiral structure as derived using \textit{HST}-ACS, VIMOS and SINFONI observations (Figures \ref{immagine}, \ref{NaDpanels}, \ref{Hapanels}, and \ref{COmaps}) overlaid to the NaD velocity field (Fig.\,\ref{NaDpanels}. The outer tip  of the  trailing arms points in the direction opposite to galactic rotation (clockwise, as indicated by the dotted arrow). The symbols $\otimes$ and $\odot$ indicate the  receding and approaching sides of the disk rotation (as traced by \Ha \, and \co). The closest part of the disk is at SW (see Sect.\, \ref{kingeom} for details). The insert with letters \textit{N} and \textit{E} indicate: the north and the east, with respect to the detector reference frame. }	      	
		    \label{model}
		\end{figure*}

		\begin{figure}
		   \centering
		    \includegraphics[scale=0.65]{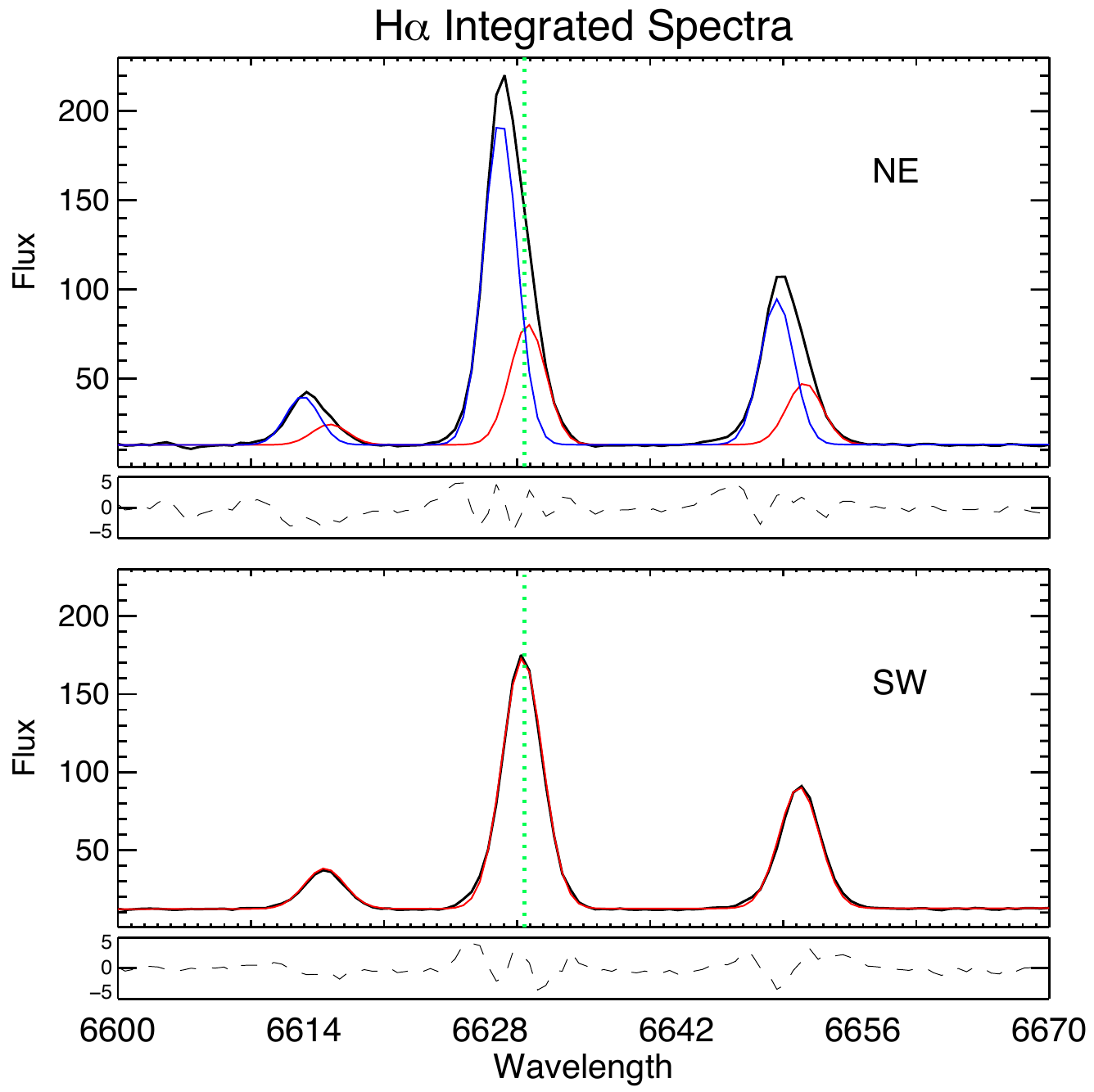}    
		     \caption{The \Ha \,  line-profiles integrated spectra (black) in two different regions of  1.2 $\times$ 1.2 kpc  along the minor axis (Fig.\,\ref{Hapanels}), respectively, towards the NE (top) and the SW (bottom).  Below, the residuals (i.e.,  data - model) are also presented. Since the asymmetry of the observed profile the NEspectrum, two components are needed to properly fit the line profile. The narrow component is shown in red in both panels, while the broad one (likely due to outflowing gas) is shown in blue only in the NE spectrum (top). The systemic velocity is shown in green with a vertical dashed line, as reference.}
		    \label{Haintegrated} 		 
		\end{figure}

In a purely rotating disk, systemic velocities are expected along the minor axis. Contrary to this, the neutral gas velocity field (Fig.\,\ref{NaDpanels}) shows strong blueshifted velocities with respect to the systemic of up to -154 $\pm$ 20 \kms \ along the NE semi-minor axis. This region of blueshifted velocities is triangular in shape, and its associated spectra indicate broad ($\sigma$\,=\,98\,$\pm$\,10 \kms), and intense (EW\,$\sim$\,3-4 \AA) profiles\footnote{The NaD-EWs associated with the outflow region should be considered upper limits, since we are unable to fit the (weak) underlying component of the NaD likely associated with the thick neutral gas disk (Sect.\ref{prop}).}.   Since the velocity dispersion of the individual neutral clouds is typically $\sim$\,15 \kms \, (\cite{SM04}), we are likely observing the combined effect of many individual clouds which, in addition of having a relatively large cloud-to-cloud relative velocity, share a common global motion. The observed neutral gas kinematics and spatial distribution of this blueshifted region are well explained by  a conical outflow  that emerges perpendicular to the disk from the nuclear region suggesting  a bipolar geometry, i.e.,  a  starburst-driven GW scenario (\cite{H87}; \cite{Rwind}). Fig.\,\ref{model} sketches the inferred geometry for the wind. In this scenario, the GW should  only be detected in NaD in the approaching cone as the receding one is not illuminated by background stars. This expectation is fully confirmed by the data, as no traces for a SW cone counterpart  are observed. The projection of the approaching cone covers $\sim$\,0.9\,$kpc^{2}$ (Fig.\,\ref{NaDpanels}) and it subtends a  narrow angle on the sky of $\sim$\,$45^{\circ}$ (as seen from the wind's origin), suggesting a collimated  structure for the wind. This opening angle agrees reasonably well with previous values found in star forming galaxies (\cite{Rwind}), and in particular to those considered to be typical in LIRGs ($\sim$\,$65^{\circ}$, \cite{R05b}). \\
From the (deprojected) velocity distribution within the wind a mean velocity of -97 \kms\, is obtained. Therefore, half of the neutral gas is travelling at typical velocities of local starburst-driven GWs in LIRGs (100-200 \kms; \cite{H00}; \cite{R05c}) reaching a mean maximum velocity (defined as the central velocity plus one half of its FWHM; \cite{R11}) of  \,-146\,\kms. The fate of the neutral outflowing gas is discussed further in Sect.\, \ref{mass}, with a comparison between the wind speeds and the galaxy's escape velocity.\\ 
There are some indications of the presence of an ionized counterpart of the neutral wind.  The observed \Ha-line profiles show the presence of a secondary kinematic component  (see Sect.\, \ref{HaLF})  preferentially in the NE sector.  This extra component is sometimes clearly seen near the border of the projection of the outflowing neutral gas cone seen in  NaD  (Fig.\,\ref{NaDpanels}).  In addition the kinematic maps of the narrow-\Ha \, component (Fig.\,\ref{Hapanels}) show some characteristics that may be related to the wind. First, while the minor kinematic axis is well aligned towards the SW, the NE counterpart (slightly) bends towards blue velocities, suggesting modest radial (e.g., outflowing) motions (Fig.\,\ref{Hapanels}). \\
To look into this feature in greater detail, we have the observed \Ha-spectra within two symmetrical regions of 1.2 $\times$ 1.2 kpc along the minor kinematic axis (black dashed boxes in Fig.\,\ref{Hapanels}). The integrated spectrum towards the NE clearly  shows a blueshifted component  (-99\,$\pm$\,5 \kms) which is not observed in the SW counterpart (Fig.\,\ref{Haintegrated}). Second,  at a radius R $\geq$ 1 kpc (PA\,=\,10$\degr$-110$\degr$) the velocity dispersion map  (Fig.\,\ref{Hapanels}\textit{c}) shows an increment  of more than 15 \kms, compared to the typical value of (45\,$\pm$\,10) \kms \, found elsewhere in  the disk (excluding the nucleus). Although the origin of these increased values is unclear, their distribution towards the NE suggests that they may be related to the wind. A consistent scenario is that the ionized gas mainly traces the outer edge of the GW rather than its inner parts, probably because the \Ha \,  emission is weighted towards the over-pressured high density gas regions. \\
The spatial information  provided by the present  IFS  data also allows us to constrain the  direction of the rotation of  the galaxy. In general, even for well-defined disks seen in projection like \obj \, (Fig.\,\ref{immagine}), the  velocity field is not sufficient to determine the direction of the rotation. In that case, if a spiral structure is observed, one has to assume if it is leading or trailing to fix the direction of rotation and, therefore, the orientation of the disk (i.e.,  its near side). However, because the cool neutral gas is identified in absorption against the stellar continuum, in the present case, one can determine that the NE edge is the far side of the disk. This  fixes the rotation direction of the disk (clockwise) and, therefore, the spiral arms observed in the \textit{HST}-NICMOS near-IR image (Fig.\,\ref{immagine}) are trailing, as illustrated in  Figure \ref{model}.\\ 
In summary, our IFS data show evidence of a neutral wind in \obj \ and have also allowed us to constrain its geometry (extension and opening angle) and kinematic properties. These suggest the presence of a conical outflow emerging from the nuclear region perpendicular to the disk at mean velocities of $\sim$\,100 \kms \,.  In addition, there is evidence of the presence of an ionized outflow that partially overlaps with the wind detected in NaD. The particular geometry of the neutral gas wind has in turn allowed us to fix the orientation and rotation direction of the disk(s).
%Steep velocity gradients along the minor axis indicate outflows with constant angular momentum, as discussed by \cite{M06}.} \\
%\textbf{Along the wind there is a large velocity gradient  ($\nabla$V(NaD)\,=\,110 $\pm$ 10  \kms $kpc^{-1}$) which suggests that an expanding bubble can efficiently accelerate the surrounding neutral gas clouds.}
 \label{kingeom}
 
%A sketch of the inferred geometry for the wind is shown in Fig.\ref{model}.

%We infer  $h_{\rm z}$\,$\sim$ 700 pc, confirming that the neutral gas is distributed in a thick disk.   \\ 
%\textit{Specifically, the height scale  (h$_z$) associated to the neutral gas is about a factor  of 4-5  larger than to that for the ionized gas, according to their  $\sigma$  /V ratios (\cite{Cresci2009}).}\\
% In this case the height scale for the neutral gas is about a factor 2-3 larger than for the stars. 
% Similar kinematic decoupling between the neutral gas and the stars have been reported by \cite{Konstantopoulos} and  \cite{Westmoquette2012} for the archetype galaxy M82. 
 
\subsection{Outflow mass rate and the ISM / IGM metal enrichment} 
%that  the  gas  is distributed in concentric spherically symmetric thin shells and 
Of particular interest is measuring the mass of gas being expelled by the wind in its cold phases (e.g., neutral) since the evacuation of this gas may be responsible for quenching  the star formation. In order to estimate the mass, the mass loss rate,  and the efficiency  of the wind, we apply a thin-shell free wind model  (FW; \cite{H00}; \cite{R02}; \cite{R05a},c), considering that the mass outflow rate and the velocities are independent of radius. The wind we observe is clearly not a thin shell, but rather a filled cone. Nevertheless, it can be considered as a series of thin shells and since we can spatially resolve it, the thin-shell formalism is still valid. We recall that  the thin-shell FW model has, however,  its own drawbacks: effects due to Rayleigh-Taylor instabilities are not taken into account when calculating the wind dynamics.\\
Some physical models (e.g., \cite{R02}; \cite{R05a}, b, c) describe in detail the outflowing material (its mass, mass rate and energy) with equations written in terms of measurable quantities, like the gas optical depth and column density.  We employ a different approach to estimate the properties (e.g., column density and mass) of the cold gas entrained in the wind. The column density is  calculated for all the spaxels within the wind using  the linear relation between the $N_{\rm HI}$ and the color excess  ($E_{\rm B-V}$), $N_{\rm HI}$\,=\,4.8\,$\times$\,$E_{\rm B-V}$\,$\times$\,$10^{21}$\,$cm^{-2}$, as  derived by \cite{Bohlin}. The mean wind column density  is  (7.7\,$\pm$\,0.4) $\times$ $10^{21}$  $cm^{-2}$, which is consistent with previous estimates for winds in dwarf starbursts and low-z LIRGs (\cite{R05c}). The color excess has been, in turn, derived directly from the present NaD-EW map (Fig.\,\ref{NaDpanels}). We made use of the relation by \cite{turatto}, $E_{\rm B-V}$\,=\,-0.04\,+\,0.51\,$\times$\,EW(NaD), derived from observations of SNe in highly reddened objects.  \cite{VKSMS}  showed that the correlation between the reddening and the strength of the NaD feature is stronger among HII galaxies like \obj \, than in LINERs and Seyfert 2 objects. In \obj, the dust embedded in the wind produces an  $E_{\rm B-V}$ in the range 0.6-2.2\footnote{\cite{Piqueras13}, derived  the 2D internal extinction structure (i.e.,  $A_{\rm V}$ map) for \obj \, using the near-IR Br$\gamma$/Br$\delta$ line ratio. However, the absence of the Br$\delta$  line in the inner part (within a radius of 0$\farcs$63)  precludes the use of the Br$\gamma$/Br$\delta$ ratio for a reliable estimation of the reddening. The different spatial distribution of the warm and neutral gas phases (see Section 4.3) adds further uncertainty when using the ratio Br$\gamma$/Br$\delta$ for estimating the E(B-V) in the outer parts of the outflow (i.e.,  neutral gas located off the disk).}. The above estimate of $N_{\rm HI}$ is very simplistic and assumes a dust-to-gas ratio observed in our galaxy ($\sim$\,150; \cite{Wilson2008}). However, in [U]LIRGs this ratio is larger (typically $\sim$\,215; \cite{Wilson2008}) and local heating (produced by high-velocity shocks) may lead to the destruction of dust grains lowering the NaD depletion level and, hence, the dust-to-gas ratio. We note that assuming solar NaD abundances (\cite{H00}) and according to the systematic uncertainties in the E(B-V)-N$_{H}$ relation (grain depletion and ionization factor),  the derived column density is uncertain to a factor of 1.5.\\
As in  \cite{Shih10}, \cite{R11}, and \cite{RV13}, we used our IFS maps to infer some physical and kinematic properties associated with the wind (e.g., spatial distribution and velocities). We first estimated the total mass ($M_{\rm w}$) of neutral gas contained in the observed  cone. The neutral gas mass in each wind's element covered by a  spaxel $k$ vary accordingly with  its  deprojected distance  from the nucleus ($R_{\rm w,k}$) and the HI column  density ($N_{\rm HI,k}$) at that location, weighted for its opening angle ($C_{\rm \Omega,k}$). For simplicity, we have restricted ourselves to a constant covering factor  (i.e.,  independent from the velocity) and assume the median value, i.e.,  $C_{\rm f}$\,=\,0.37, observed in LIRGs by \cite{R05c}. Therefore the wind mass is:

\begin{center}
	 \begin{equation} 
{M}_{\rm w}\,=\, 5.6 \times 10^{8} \sum \limits_{\rm k=1}^N \left(\frac{C_{\rm \Omega,k}}{0.4} C_{\rm f}\right) \left(\frac{R_{\rm w,k}^{2}}{100 \, kpc^{2}}\right)   \left(\frac{N_{\rm H,k}}{10^{21} cm^{-2}}\right) \ M_{\rm \sun}.
\label{masswind}
	\end{equation}
  \end{center}
  The amount of the neutral gas mass expelled by the wind in \obj\ is  $M_{\rm w}$\,=\,(8.5  $\pm$ 0.9) in log\,[M/$M_{\rm \sun}$] units, which is within the typical values found in [U]LIRGs (\cite{Rwind}). We note that according to the systematic uncertainties in the calculation of the column density,  $M_{\rm w}$ is uncertain to at least a factor of 1.5. As GWs are typically bi-conical in shape, and we are probing only the approaching cone. Assuming a similar ISM distribution on both sides of the galaxy a better estimate to the total mass expelled by the wind would be $M_{\rm w}$\,$\sim$\,8.8 log\,[M/$M_{\rm \sun}$] (a factor two higher than our estimation). The total mass\footnote{We considered the total mass associated with double-horned H{\, \small I}  profile, i.e.,  the mass within the primary beam of the single-dish Green Bank Telescope observations  (\cite{cou11}).} of neutral gas in \obj, as determined by single-dish radio observations of the H$\small{I}$ 21cm line (\cite{cou11}), is M$_{\rm HI}\,= \,2.7 \times 10^{9}$ M$_{\rm \odot}$. Therefore, at least $\sim$\,10$\%$ of the neutral gas in F11506-3851 is expected to be involved in the wind (which is likely a conservative estimate, given that the NaD wind can only be traced against regions with a strong background continuum).\\
We  then computed the time-averaged outflow rate ($\dot{M_{\rm w}}$) following \cite{R05c}: 

\begin{center}
 \footnotesize{ 
 \begin{equation} 
 		\dot{M}_{\rm w}\,=\, 11.5 \times \sum \limits_{\rm k=1}^N \left(\frac{C_{\rm \Omega,k}}{0.4} C_{\rm f}\right) \left(\frac{R_{\rm w,k}}{10  \, kpc}\right)  \left(\frac{N_{\rm H,k}}{10^{21} cm^{-2}}\right) \left(\frac{V_{\rm w,k}}{200  kms^{-1}}\right) \ M_{\rm \sun}/yr.
	\end{equation}}
  \end{center} 
The notation follows that of Eq. (\ref{masswind}), and $V_{\rm w,k}$ corresponds to the deprojected velocity in each spaxel. We permitted every spaxel have its own speeds, since small velocity differences  are consistent with projection effects. However, the deviations with respect to a wind with constant velocity (i.e.,  the median wind speed) are small ($<$\,15$\%$). We estimate that the effects on $\dot{M_{\rm w}}$  due to the average stellar contribution to the NaD profile\footnote{In order to estimate this effect  the spectra over the outflowing region were \lq\lq decontaminated\rq\rq \ by subtracting the stellar contribution obtained from the continuum fitting to the integrated spectrum (Sect.\,\ref{stellarNaD})  at the spaxel \Ha \ velocity. \Ha \  was preferred to CO as a proxy of the stellar velocity, as it covers a larger FoV, and it was obtained from the same  data cube as NaD. Furthermore, the \Ha \ velocity field shows similar overall characteristics as the stellar as presented in Figures \ref{Hapanels} and \ref{COmaps}.} over the outflowing region are small ($<$\,20$\%$). We infer a mass loss rate of $\dot{M_{\rm w}}$\,=\,48 $M_{\rm \sun}/yr$. The relevance of the GW feedback is quantified using the mass loading parameter  ($\eta$) defined  as the mass outflow rate normalized to the corresponding SFR. For  \obj\   we obtained  $\eta$\,=\, 1.4, indicating that the wind contributes significantly to the exhaustion of the gas reservoir available for the star formation. A more detailed analysis considering the actual spatial distribution of SF and the characteristics of the wind inferred from the present IFS data shows a more complex picture, as discussed in Sect.\,\ref{origin}. \\
A common way to determine if winds can  pollute the IGM is to compare the wind speed with the local escape velocity derived from a gravitational model for the host galaxy. Considering a truncated isothermal sphere, the escape velocity  ($v_{\rm esc}$) is related to the rotation speed ($v_{\rm rot}$) at a radius \textit{r} as
 \begin{center}
	 \begin{equation} 
 		v_{\rm esc}(r)\,=\,\sqrt{2 \times v_{\rm rot}^{2} \times \left[1 + ln \left(\frac{R_{\rm max}}{r}\right)\right]} 
	\end{equation}
  \end{center}
  where $R_{\rm max}$ is the truncation radius. We adopt  the deprojected stellar rotation velocity  (i.e.,   235 \kms) as the rotation velocity, Ê since  the  stellar dynamics in galaxies represent a  sensitive probe of the galactic gravitational potential and $R_{\rm max}$/\textit{r}\,=\,10\, following \cite{R02}, inferring a $v_{\rm esc}$ of $\sim$\,430 \kms.  Since the inclination corrected speeds for the neutral gas clouds driven by the wind range  within 30\,-\,200\,\kms\,  most (all) of the outflowing material will not be able to escape the gravitational potential of the host galaxy. Therefore, although the mass loading factor is about 1.4, the outflowing material will not contribute to the chemical enrichment of the surrounding IGM. \\
Given that most of the gas entrained in the wind will remain gravitationally bound to the host galaxy, but not virialized, it is expected that this material will fall back onto the galactic disk. This cycle (i.e.,  the expulsion and the rain-back of gas) could be responsible for the assembly of an extra-planar gas reservoir.  This process could be somewhat analogous to that observed with deep observations of H{\small I} 21cm line emission in several nearby disk galaxies, where neutral hydrogen gas is deposited into the galaxy's halo environment mainly by stellar feedback (in the most extreme cases out to a few tens of kpc above the main galaxy disk;  e.g.,  \cite{fra02};  \cite{Fraternali2005}; \cite{boo05}; \cite{oos07}; \cite{fra08}). This extra-planar neutral gas is characterized by slower rotation compared to the  galaxy disk (most pronounced at small galactic radii), it can have an asymmetric distribution, and it may contain up to $\sim$\,10$^{9}$ M$_{\odot}$ of neutral gas (in some cases up to $\sim$\,30$\%$ of the galaxy's total H{\small I} content; see \cite{oos07}; \cite{boo08} for examples). \\
Because the wind in \obj \ originates from the nuclear region it may already have a relatively low angular momentum compared to the gas in the ionized thin disk. If NaD can survive the process of expulsion and re-accretion it may perhaps explain our intriguing result that the main NaD component in \obj \, traces a slowly rotating, somewhat asymmetric thick gas disk (Fig.\,\ref{DVP}). Geometrical effects or an over-abundance of NaD in the extra-planar gas may perhaps favor the detection of this slowly rotating thick disk above a potential neutral NaD counterpart in the thin disk (see also Sect. \ref{prop}.)

 \label{mass}

\subsection{The origin of the wind. Witnessing the quenching of the nuclear star formation process?} 

As already shown in previous sections, the outflow in \obj \, is mostly detected in the neutral gas component of the interstellar medium with a weak indication of the ionized gas counterpart. In addition to this, the apex of the cone-like morphology has been identified with the center (both kinematic and photometric center) of the galaxy.  However, while most neutral and ionized gas outflows in [U]LIRGs have been associated with either an intense star formation, an AGN, or both, the center of \obj \, does not appear to be undergoing a luminous active phase. On the one hand, the AGN appears to be weak with a low contribution ($<$\,4\%) to the total luminosity, as estimated from Spitzer/MIPS 24$\mu$m measurements (\cite{PereiraSantaella2010}). On the other hand, the nuclear region is a weak source of ionized gas as traced by both the H$\alpha$ (Fig.\,\ref{Hapanels} and \cite{RZ10}) and Br$\gamma$ 2.166 $\mu$m (\cite{Piqueras12}) emission lines. Specifically, only less than 3 percent of the observed Br$\gamma$ emission (i.e.,  SFR) is found in the central 400 x 400 pc region (\cite{Piqueras12}). In addition, the \Ha \,  and the Br$\gamma$  maps trace a circumnuclear ring of luminous star forming clumps, and a nucleus that appears to be depleted of current star formation. This is supported by the relatively strong  [FeII] emission (Br$\gamma$/[FeII]1.64$\mu$m\,$\sim$\,1) indicating the presence of supernovae, and therefore of a recent nuclear starburst with an age in the range of about 6.7-7.5 Myr, according to stellar population models for instantaneous bursts (\cite{Leitherer1999}). \\
In order to check the SNe-scenario as origin of the GW observed in \obj, we compare the expected energy budget released by the SNe within the wind lifetime ($\sim$\,10 Myr \footnote{The wind lifetime has been inferred as the ratio between  the maximum extent and the mean velocity of the wind (1.2 kpc and 95 \kms, respectively.}) and that of the wind. To predict the nuclear SNe rate we use the [FeII]-flux measurements (within the inner 400 $\times$ 400 pc, \cite{Piqueras13}) corrected for the internal extinction\footnote{We express the near-IR internal extinction ($A_{\rm [FeII]}$) in terms of the visual extinction $A_{\rm v}$ (i.e.,  $A_{\rm [FeII]}$\,=\,0.1961$A_{\rm v}$) by using the extinction law described in \cite{Calzetti2000}. In absence of a direct nuclear $A_{\rm v}$-measurement, we consider the mean value of 7.8 magnitudes derived by \cite{Piqueras13}, which is likely a lower limit according to the radial profile measured by these authors).}. We infer that the nuclear SNe rate is about 2.0 $\times$ $10^{-3}$ $yr^{-1}$ assuming the [FeII] flux to SNe rate expression of  \cite{AH03}, derived from the  study of M82 and NGC 253 (see \cite{Bedregal2009} for other derivations of SNe rates based on [FeII]). Since the internal extinction could be underestimated, an upper limit to the SNe rate can be derived from the high angular resolution radio measurements (\cite{Baan2006}). For a nuclear luminosity of 8.34 $\times 10^{21}$ W Hz$^{-1}$ at 4.8 GHz, a SNe rate of about 0.2 yr$^{-1}$ is derived applying relations based on M82 radio supernovae (\cite{Huang1994}), and evolutionary models of the radio emission in starbursts (\cite{PerezOlea1995}). 
Assuming an energy budget of $10^{51}$ erg for each SNe-event, the total energy injected within the lifetime of the wind is 55.3$-$57.3 in log[E/$erg^{-1}$] units. The energy of the wind ($E_{\rm w}$) is inferred as the sum of the kinetic energy,  $v_{\rm k}^{2}/2$, and a term, i.e.,  $3\sigma_{\rm k}^{2}/2$, associated with turbulences for all  spaxels $\textit{k}$ where the outflow has been detected,  (see \cite{R05c}):  

\begin{center}
 \footnotesize{ 
 \begin{equation} 
 		E_{\rm w}\,=  \sum \limits_{\rm k=1}^N M_{\rm k}\, \times \left(\frac{v_{\rm k}^{2}}{2}+\frac{3\sigma_{\rm k}^{2}}{2}\right).
	\end{equation}}
  \end{center} 
  
We derived a  total energy of the wind of 56.4 in units of log[E/$erg^{-1}$], which lies within the range of the total energy injected by the supernovae as derived above. We have previously derived (see Sect.\,\ref{mass})  an outflowing mass rate in neutral gas of about 48 M$_{\rm \odot}$ yr$^{-1}$, close to the star formation rate (i.e.,  34 M$_{\rm \odot}$ yr$^{-1}$; Table\,\ref{source}) derived from the infrared luminosity. Therefore, the overall mass loading ($\eta$) factor (i.e.,  ratio of outflowing  gas mass to total star formation rate) is about 1.4. However, from our VIMOS and SINFONI data it is clear that while the current  star formation covers an area of several kpc$^{2}$ and is distributed in several clumps, the neutral gas outflowing material is confined to a much smaller ($<$\,1 $kpc^{2}$) and well-defined region that appears to be associated with the nucleus, where no intense star formation is underway. Since the global mass loading factor of the galaxy is close to one (i.e.,  1.4), the corresponding local mass loading factor for the nucleus and its associated star formation would be much larger than one (i.e.,  $\eta$\,$>>$\,1) indicating negative-feedback and therefore the possibility of seeing the  quenching of the star formation as a result of the combined supernovae and stellar winds.  The relatively strong [FeII] (i.e.,  presence of stellar populations several Myrs old where massive stars have exploited as SNe) and the weak Br$\gamma$ (i.e.,  lack of massive young stellar populations few Myrs old) fluxes, as well as the presence of the wind, suggests that we are witnessing a quenching event in \obj.\\
While the scenario above can be qualitatively valid and needs further exploration with higher angular resolution, the overall  scenario suggested by our data is far more complex. The nucleus is the strongest warm molecular gas emitting region (Piqueras-Lopez et al. 2012) contributing 21$\%$ of the total, as traced by the H$_{\rm 2}$ 1$-0$S(1) emission line. The mass of warm molecular gas in the nuclear region can be computed following Dale et al. (2005) as M$_{\rm H2}^{warm}\,=\,5.08 \times F(H2) \times D_L^2$ where F(H2) is the extinction corrected 2.122 $\mu$m H$_2$ line flux in units of 10$^{-16} W m^{-2}$ and D$_L$ is the luminosity distance in Mpc. For an observed nuclear flux of 1.7 $\times 10^{-18} W m^{-2}$ (\cite{Piqueras12}), and an extinction corrected flux of 6.8 $\times 10^{-18} W m^{-2}$, the amount of warm molecular gas in the nuclear region is $\sim$\,750 M$_{\rm \odot}$ corresponding  to a total mass in the range 0.75-75 $\times$ $10^{8}$ M$_{\rm \odot}$ in cold molecular gas if the range of warm to cold molecular gas ratios observed in starburst galaxies is assumed (\cite{Dale2005}). Thus, assuming an efficiency of 10\% in the conversion of cold gas into stars, there is enough molecular gas in the nuclear region to form between about 10$^7$ M$_{\rm \odot}$ and 10$^8$  M$_{\rm \odot}$ of new stars in future (recurrent) episodes of star formation (as seen for a small sample of nearby spirals   by \cite{Gallagher13}), even if the nucleus is currently  undergoing a quenching phase due to the previous starburst that took place several million years ago.

 \label{origin}

\section{Summary and conclusions}

On the basis of optical (VIMOS) and near-IR (SINFONI)  IFS data we have studied the kinematic and dynamical properties of the stellar component and gas phases (ionized and neutral) in the luminous infraRed Galaxy  \obj \,  (ESO 320-G030) using as tracers the NaD absorption doublet, H$\alpha$, and the CO bands.  The conclusions of the present study can be summarized as follows:

\begin{enumerate}

\item  --- \textit{Kinematics and dynamical support for the stars and gas phases (warm and neutral). } The ionized ISM is in a rather turbulent state as indicated by its relatively high velocity dispersion ($\sigma$\,$\sim$\,45 \kms), which is intermediate between  low-luminosity normal spirals  and high-z SFGs.  Despite this, the ionized gas is largely dominated by rotation and it is distributed in a dynamically cold disk as indicated by its large V/$\sigma$  ratio ($\sim$\,4.5).  The stellar velocity field is also dominated by ordered large-scale rotational motions, although it slightly lags in velocity compared to e the ionized gas.  The stars have also a larger random velocity component and, therefore,  the V/$\sigma$ ratio (2.4)  suggests that the stellar disk has lower rotational support and a larger height scale  than the ionized gas  (125 vs. 35 pc).  In contrast to what  is found for the  stars and the warm ionized gas,  the neutral gas shows complex kinematics. Excluding a region affected by strong radial motions (see below), the velocity field of the main NaD component  indicates a modest large-scale rotation can be interpreted as due to a dynamically hot  slowly rotating disk mainly supported by random motions (V/$\sigma$\,$\sim$1), with a height scale 
larger than the ionized gas and stars of about a factor of 5 and 2, respectively.  These results are evidence of a complex stratification and dynamics of the different galaxy components in  \obj.  

\item  --- \textit{Stellar and ISM contributions to the  NaD feature.}  Applying a spaxel-by-spaxel  spectral fitting decomposition to the NaD profiles we have been able to disentangle two distinct components for which their 2D structural and kinematic properties have been inferred.  One of these components (main) is likely tracing the neutral gas clouds of the ISM, as its EW-map is unrelated to that for the stellar continuum and it shows strong absorptions.  The secondary component has properties fully consistent with a stellar origin, as can be seen by kinematic properties similar to those obtained from the infrared CO bands, and by its low EW values. These results are also supported by the application of a stellar-continuum modeling approach to the integrated spectrum, which indicates that most of the NaD absorption ($\sim$\,2/3) is interstellar in origin, while only a small contribution is due to the stars (1/3).          

\item  --- \textit{Evidence of a neutral outflow.}  The velocity field of the neutral gas shows strong blue-shifted velocities along one semi-minor axis (towards the NE), which indicate strong radial motions emerging from the nuclear region perpendicular to the disk, and it is, therefore,  naturally interpreted in the starburst driven galactic wind scenario. The triangular morphology of this region, and its kinematic properties (e.g., velocities of up to $\sim$\,200\,\kms), are also consistent with a conical outflow (opening angle $\sim$\,45 degrees) extending at least 1.8 kpc. Although the presence of an ionized outflow counterpart is less obvious, there is evidence of the presence of radial motions of ionized gas in a region that overlaps with that of the neutral wind. The particular geometry of the neutral gas has in turn allowed us to fix the orientation and rotation direction of the galaxy disk(s).  

\item  --- \textit{Wind feedback and metal enrichment.} Under a free wind model (e.g., Heckman et al. 2000), we have inferred an outflowing mass rate of about 48 M$_{\rm \odot}$\,yr$^{-1}$, which considering the total SFR of this galaxy implies a global mass loading factor ($\eta$\,=\,M$_{\rm w}$/SFR) of 1.4. However, on the basis of the measured outflowing velocities and the dynamical mass (i.e.,  escape velocity) of the galaxy, it is unlikely that a significant fraction of the outflowing material would escape the galaxy to chemically enrich the surrounding IGM. Therefore, the outflowing material will fall back onto the galaxy disk, which would explain the lagging and the thickening of the neutral gas disk.

\item  --- \textit{Wind feedback and nuclear SF quenching.} The nuclear region has been identified as the site where the outflow originates. The large local mass loading factor ($>>$\,1) associated with this region, together with its relatively modest ongoing SF (seen via Br$\gamma$ and H$\alpha$ emissions) and the evidence of SNe  (i.e.,  [FeII] emission) from a $\sim$\,7 Myr  old nuclear burst, suggest that the outflow has been generated by the effects of these SNe and it  has effectively quenched star formation over the past few Myr in this region. However, the large nuclear reservoir of cold molecular gas, as inferred from the H$_{\rm 2}$ 1$-0$S(1) emission, suggests that recurrent episodes of SF may be occur again in the future.

%textbf{with a constant angular momentum, ERASE!}
  
\end{enumerate}

\label{concl}

\begin{acknowledgements}

%We thanks the anonymous referee for helpful comment which greatly improved the discussion and the content of this paper. 
The authors wish to thank the referee  (C.Tremonti) for the detailed and constructive report. Thanks are also due to Jorge Jimen\'ez-Vicente  for the useful discussions. Sara Cazzoli is supported through a  European Community Marie Curie Fellowship. This work was funded in part by the Marie Curie Initial Training Network ELIXIR of the European Commission under contract PITN-GA-2008-214227. This work has been supported by the Spanish Ministry of Science and Innovation (MICINN) under grants  AYS2010 and AYA2012. This work is based on observations carried out at the European Southern observatory, Paranal (Chile), Programs 076.B-0479(A), 078.B-0072(A) and 081.B-0108(A). This research has made use of the NASA/IPAC Extragalactic Database (NED) which is operated by the Jet Propulsion Laboratory, California Institute of Technology, under contract with the National Aeronautics and Space Administration. 

%ESP2007-65475-C02-01, AYA2010-21161-C02-01, AYA2012-32295, AYA2012-39408-C02-C01.
 %This paper made use of the plotting package \textit{jmaplot}, developed by Jes\'us Ma\'iz-Appelaniz \cite{Maiz04}.

\end{acknowledgements}

%   \item Title -- change for something like: Multi wavelength IFS of IRAS F11506-385: the role of SNe-driven galactic wind in the quenching of star formation

\end{document}